\documentclass[cernpreprint,USenglish]{na61doc}
\usepackage{amsmath}
\usepackage{url}
\usepackage{multirow}
\usepackage{makecell}

\usepackage{colortbl}
\definecolor{darkred}{rgb}{0.5,0,0}
\definecolor{darkblue}{rgb}{0,0,0.5}
\definecolor{firebrick}{rgb}{0.75,0.125,0.125}
\definecolor{darkgreen}{rgb}{0,0.5,0}

\usepackage[colorlinks=true,linkcolor=firebrick,citecolor=darkgreen,urlcolor=darkblue]{hyperref}

\usepackage{xspace}
\usepackage{enumerate}
\ShineTitle{$K^{*}(892)^0$ meson production in inelastic p+p interactions at 158~\GeVc beam momentum measured by \NASixtyOne at the CERN SPS}

\PreprintIdNumber{CERN-EP-2020-002}

\ShineJournal{}	

\ShineJournalRef{published in Eur. Phys. J. C80, 460 (2020)}
\ShineDOI{10.1140/epjc/s10052-020-7955-1}

\ShineAbstract{
The measurement of $K^{*}(892)^0$ resonance production via its $K^{+}\pi^{-}$ decay mode in inelastic p+p collisions at beam momentum 158~\GeVc ($\sqrt{s_{NN}}=17.3$~\GeV) is presented. The data were recorded by the \NASixtyOne hadron spectrometer at the CERN Super Proton Synchrotron. The \textit{template} method was used to extract the $K^{*}(892)^0$ signal and  double-differential transverse momentum and rapidity spectra were obtained. The full phase-space mean multiplicity of $K^{*}(892)^0$ mesons was found to be 
 $(78.44 \pm 0.38 \mathrm{(stat)} \pm 6.0 \mathrm{(sys)) \cdot 10^{-3}}$. 
The \NASixtyOne results are compared with the \EposLong and Hadron Resonance Gas models as well as with world data from p+p and nucleus-nucleus collisions.  
}

\newcommand{\eV}{\ensuremath{\mbox{e\kern-0.1em V}}\xspace}
\newcommand{\GeV}{\ensuremath{\mbox{Ge\kern-0.1em V}}\xspace}
\newcommand{\MeV}{\ensuremath{\mbox{Me\kern-0.1em V}}\xspace}
\newcommand{\GeVc}{\ensuremath{\mbox{Ge\kern-0.1em V}\!/\!c}\xspace}
\newcommand{\GeVcc}{\ensuremath{\mbox{Ge\kern-0.1em V}\!/\!c^2}\xspace}
\newcommand{\AGeV}{\ensuremath{A\,\mbox{Ge\kern-0.1em V}}\xspace}
\newcommand{\AGeVc}{\ensuremath{A\,\mbox{Ge\kern-0.1em V}\!/\!c}\xspace}
\newcommand{\MeVc}{\ensuremath{\mbox{Me\kern-0.1em V}/c}\xspace}

\newcommand{\cm}{\ensuremath{\mbox{cm}}\xspace}

\newcommand{\s}{\ensuremath{\mbox{s}}\xspace}

\newcommand{\dd}{\ensuremath{{\text{d}}}\xspace}
\newcommand{\dedx}{\ensuremath{\dd E\!/\!\dd x}\xspace}
\newcommand{\tof}{\ensuremath{\textup{\emph{tof}}}\xspace}
\newcommand{\pt}{\ensuremath{p_{\text{T}}}\xspace}

\newcommand{\GeantThree}{{\scshape Geant3}\xspace}

\newcommand{\EposLong}{{\scshape Epos1.99}\xspace}

\newcommand{\CernVM}{\textsc{Cern\-\kern-0.05emVM}\xspace}

\begin{document}

\maketitle

%\linenumbers

\section{Introduction and motivation}\label{sec:intro}

Strange hadron production is believed to be an important tool to study the dynamics of high-energy collisions. In collisions achieving high energy densities strangeness production was predicted to be enhanced~\cite{Rafelski:1982pu} as a result of the decrease of the mass of strangeness carriers due to partial chiral symmetry restoration. The $K^{*}(892)^0$ resonance state contains an $\bar{s}$ valence quark and is therefore sensitive to the level of strangeness production. Thus, the data on $K^{*}(892)^0$ meson production provide a more complete understanding of hadron chemistry.

Measurements of the production of short-lived resonances are a unique tool to understand the less known aspects of high energy collisions, especially their time evolution. In heavy ion collisions the yields of resonances may help to distinguish between two possible freeze-out scenarios: the sudden and the gradual one~\cite{Markert:2002rw}. Namely, the ratio of $K^{*}(892)^0$ to charged kaon production may allow to estimate the time interval between chemical (end of inelastic collisions) and kinetic (end of elastic collisions) freeze-out. The lifetime of the $K^{*}(892)^0$ resonance ($\approx 4$ fm/$c$) is comparable to the expected duration of the rescattering hadronic gas phase between the two freeze-out stages.
Consequently, a certain fraction of $K^{*}(892)^0$ resonances will decay inside the fireball. The momenta of their decay products are expected to be significantly modified by elastic scatterings, preventing the experimental reconstruction of the resonance via an invariant mass analysis. In such a case a suppression of the observed $K^{*}(892)^0$ yield is expected.
Such an effect was indeed observed in nucleus-nucleus collisions at Super Proton Synchrotron (SPS) and Relativistic Heavy Ion Collider (RHIC) energies~\cite{Blume:2011np, Anticic:2011zr, 
Adams:2004ep, Abelev:2008yz, Aggarwal:2010mt, Nasim:2019ysf}. The ratio of $K^{*}/K$ production ($K^{*}$ stands for $K^{*}(892)^0$, $\overline{K^{*}}(892)^0$ or $K^{*\pm}$, and $K$ denotes $K^{+}$ or $K^{-}$) showed a decrease with increasing system size as expected due to the increasing rescattering time between chemical and kinetic freeze-out. 
The same effect was recently reported also by the ALICE Collaboration at the Large Hadron Collider (LHC)~\cite{Abelev:2014uua, Adam:2017zbf, Albuquerque:2018kyy, Acharya:2019qge}. 

When looking at the energy dependence of the $K^{*0}/K^{-}$ ratio\footnote{In ALICE at LHC and STAR at RHIC papers. e.g. Refs.~\cite{Adam:2017zbf, Acharya:2019qge, Aggarwal:2010mt}, the results for $K^{*}(892)^0$ and $\overline{K^{*}}(892)^0$ were combined and averaged and denoted by the symbol $K^{*0}$; the ratios were measured at mid-rapidity.}
in central Pb+Pb or Au+Au collisions, a bit larger suppression of $K^{*0}$ is observed for the 2.76~TeV\footnote{The $K^{*0}/K^{-}$ ratios in Pb+Pb collisions at $\sqrt{s_{NN}}$ = 2.76~TeV and 5.02~TeV are in agreement within uncertainties~\cite{Acharya:2019qge}.} LHC energy~\cite{Adam:2017zbf} when compared to the top RHIC ($\sqrt{s_{NN}}$=200~\GeV) energy~\cite{Aggarwal:2010mt}, namely $K^{*0}/K^{-}$ = $0.180 \pm 0.027$ ($0.186 \pm 0.027$) for the 0--5\% (5--10\%) central Pb+Pb reactions at LHC and $0.20 \pm 0.04$ for the 0--10\% most central Au+Au interactions at RHIC. Those values can be compared with those for p+p interactions, which are $0.307 \pm 0.043$ at LHC~\cite{Adam:2017zbf} and $0.34 \pm 0.05$ at RHIC~\cite{Aggarwal:2010mt}. Thus, the $K^{*0}/K^{-}$ ratio in central Pb+Pb collisions at LHC (2.76~TeV) drops to 59 (61)\% of the value found for p+p interactions. For RHIC energies this drop is similar and equals 59\%.

In the NA49 experiment at the CERN SPS $K^{*}(892)^0$ and $\overline{K^{*}}(892)^0$ meson production was analyzed separately and the corresponding (almost $4\pi$) mean multiplicities obtained in the 23.5\% most central Pb+Pb collisions at $\sqrt{s_{NN}}$=17.3~\GeV are $10.3 \pm 2.5$ and $5.2 \pm 1.7$, respectively~\cite{Anticic:2011zr}. They can be rescaled (using the mean number of wounded nucleons; factor 362/262, see also Table~\ref{tab:Kstar_Kp_Km}) to the 5\% most central collisions, resulting in mean multiplicities of $14.2 \pm 3.5$ and $7.2 \pm 2.3$, respectively. Their average, divided by the $\langle K^{-} \rangle$ multiplicity ($51.9 \pm 3.6$) for the 5\% most central Pb+Pb collisions~\cite{Afanasiev:2002mx} results in the ratio $0.5 \cdot (\langle K^{*}(892)^0 \rangle + \langle \overline{K^{*}}(892)^0 \rangle)/ \langle K^{-} \rangle$ of $0.21 \pm 0.04$ which is similar to the value 
$K^{*0}/K^{-} = 0.20 \pm 0.04$ measured in the 10\% most central Au+Au collisions at RHIC~\cite{Aggarwal:2010mt}. 
Finally, the ratio $0.5 \cdot (\langle K^{*}(892)^0 \rangle + \langle \overline{K^{*}}(892)^0 \rangle)/ \langle K^{-} \rangle$ for p+p interactions at the same SPS energy can be estimated as $0.48 \pm 0.04$~\cite{Anticic:2011zr, Aduszkiewicz:2017sei}. 
Thus, at SPS energy the resonance to non-resonance ratio in central Pb+Pb drops to about 43--44\% of the value for p+p interactions. This effect is even stronger than the one observed at RHIC and LHC and might suggest that the lifetime (calculated in the $K^{*0}$ rest frame; see Eq.~(\ref{eq:time_freezeout}) in Sec.~\ref{sec:comparison_time}) of the hadron gas system created in central nucleus-nucleus collisions at the SPS is longer than that at higher energies. Eventually, resonance regeneration processes start to play a role for higher energies counteracting the $K^{*0}$ suppression due to rescattering. It should also be pointed out that the whole picture assumes that the conditions at chemical freeze-out of p+p and Pb+Pb collisions are the same. More detailed calculations of the time between freeze-outs, both in the $K^{*0}$ rest frame and in collision center-of-mass reference system, are given in Sec.~\ref{sec:comparison_time}.

The results for p+p collisions provide an important base-line for heavier nucleus-nucleus systems. So far the $K^{*0}/K^{-}$ ratio for p+p interactions did not show large differences between the top RHIC and four LHC energies~\cite{Abelev:2012hy, Adam:2017zbf, Acharya:2019wyb, Acharya:2019qge}. Most of the results at lower energies are less reliable due to large uncertainties, see the compilation in Ref.~\cite{Abelev:2012hy}, and new points in Refs.~\cite{Anticic:2011zr, Adam:2017zbf, Acharya:2019wyb, Acharya:2019qge}.  
This emphasizes the need to obtain high precision p+p data at energies lower than the top RHIC energy. Continuing considerations for p+p collisions, a very intriguing effect was reported in the most recent ALICE analysis of the multiplicity dependence in p+p collisions~\cite{Dash:2018cjh, Acharya:2019bli}. The $K^{*0}/\langle K^{\pm} \rangle$ and $K^{*0}/K^0_S$ ratios decrease when going from low-multiplicity to high-multiplicity p+p interactions at the LHC energies. This may be an indication of a hadronic phase with significant non-zero lifetime even in p+p collisions. 
 
The transverse mass spectra and yields of $K^{*}(892)^0$ mesons are also important inputs for Blast-Wave models (determining kinetic freeze-out temperature and transverse flow velocity) and Hadron Resonance Gas models (determining chemical freeze-out temperature, baryochemical potential, strangeness under-saturation factor, system volume, etc.). Those models significantly contribute to our understanding of the phase diagram of strongly interacting matter. In principle, the precise determination of transverse flow velocity is attractive due to the fact, that recent LHC, RHIC and even SPS results suggest that dense and collectively behaving system may appear also in collisions of small nuclei, or even in elementary interactions. Finally, the study of resonances in elementary interactions contributes to the understanding of hadron production, due to the fact that products of resonance decays represent a large fraction of the final state particles. Resonance spectra and yields provide an important reference for tuning Monte Carlo string-hadronic models.

The study of $K^{*}(892)^0$ and/or $\overline{K^{*}}(892)^0$ production in p+p collisions at RHIC energies was performed by the STAR~\cite{Adams:2004ep} and PHENIX~\cite{Adare:2014eyu} experiments and at LHC energies by ALICE~\cite{Abelev:2012hy, Adam:2017zbf, Acharya:2019wyb, Acharya:2019qge, Acharya:2018orn, Acharya:2019bli}. The NA49 experiment performed the measurements in inelastic p+p collisions at beam momentum  of 158~\GeVc (CERN SPS)~\cite{Anticic:2011zr}. Also the LEBC-EHS facility at the CERN SPS measured $K^{*}(892)^0$ and $\overline{K^{*}}(892)^0$ production in p+p collisions at 400~\GeVc~\cite{AguilarBenitez:1991yy}. Finally, results obtained at the energies of the CERN Intersecting Storage Rings (ISR) were published in Refs.~\cite{Drijard:1981ab, Akesson:1982jg}.

This paper reports measurements of $K^{*}(892)^0$ resonance production via its $K^{+}\pi^{-}$ decay mode in inelastic p+p collisions at beam momentum of 158~\GeVc ($\sqrt{s_{NN}}=17.3$~\GeV)\footnote{The analysis of $\overline{K^{*}}(892)^0$ as well as $K^{*}(892)^0$ and $\overline{K^{*}}(892)^0$ at lower SPS energies is a subject of future \NASixtyOne paper.}~\cite{AT_PhD}. The data were recorded by the \NASixtyOne hadron spectrometer~\cite{Abgrall:2014xwa} at the CERN SPS. Unlike in the previous NA49 analysis~\cite{Anticic:2011zr} at the same beam momentum, the \textit{template} method was used to extract the $K^{*}(892)^0$ signal. This method was found to allow a more precise background subtraction than the \textit {standard} procedure. Moreover, the large statistics \NASixtyOne data (about 52.5M events recorded with the interaction trigger compared to 2.5M p+p events analysed in NA49~\cite{Claudia_priv, Claudia_H_PHD}) 
allowed to obtain high quality double-differential transverse momentum and rapidity spectra of $K^{*}(892)^0$ mesons. The paper is organized as follows. Section~\ref{sec:setup} briefly describes the \NASixtyOne detector. Section~\ref{sec:method} discusses the analysis procedures, including event and track cuts, method of signal extraction, corrections, and evaluation of uncertainties. The final results are presented in Section~\ref{sec:results} and their comparison with world data and models 
in Section~\ref{sec:comparison}. A summary Section~\ref{sec:summary} closes the paper.

%%%
%%% Local Variables: 
%%% mode: latex
%%% TeX-master: "main"
%%% End: 
    
\section{Experimental setup}\label{sec:setup}

\begin{figure*}
  \centering
  \includegraphics[width=0.8\textwidth]{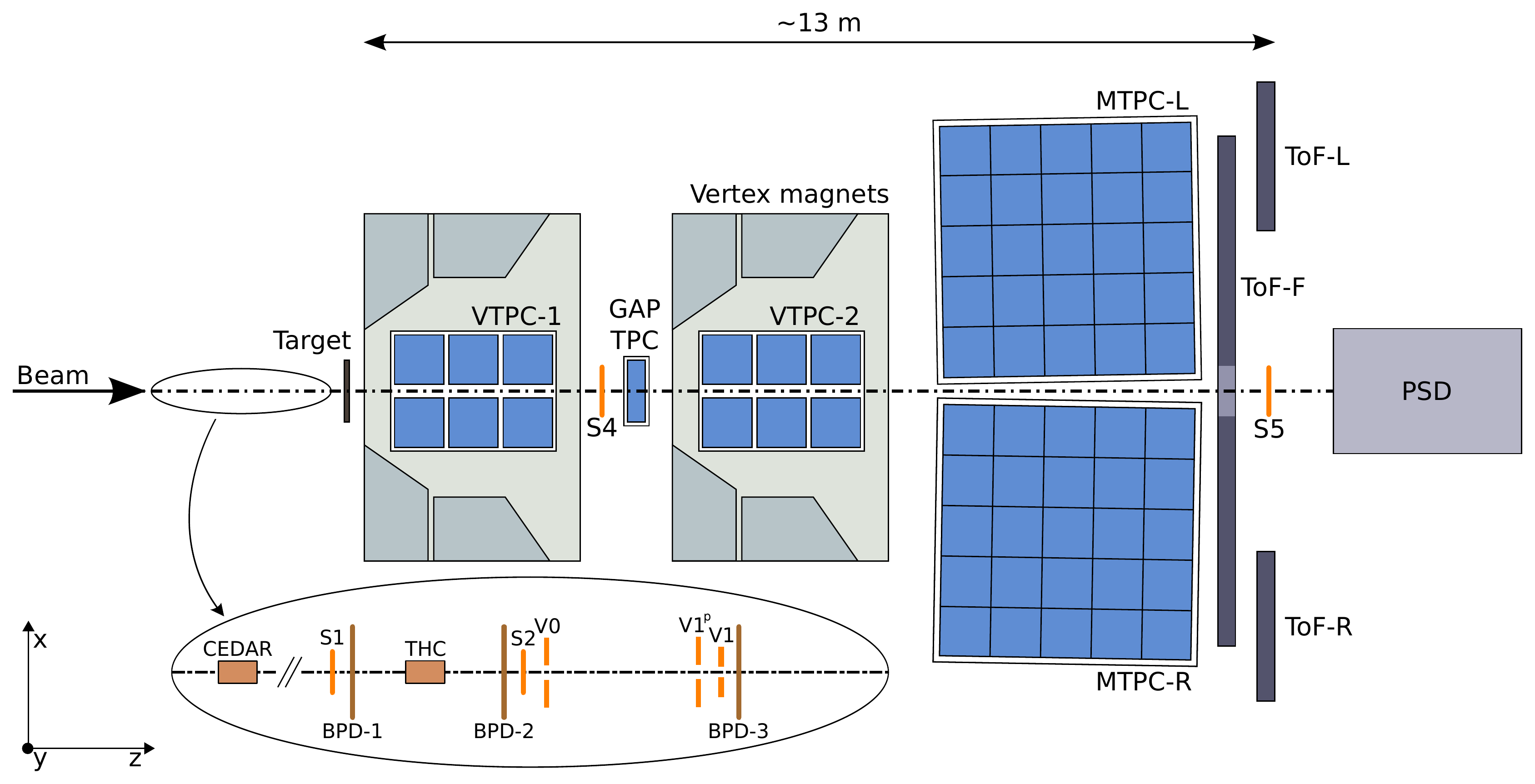}
  \caption[]{
    (Color online) The schematic layout of the NA61/SHINE experiment at the CERN SPS (horizontal cut, not to scale).
The beam and trigger detector configuration used for data taking in 2009 is shown in the inset (see Refs.~\cite{Abgrall:2013qoa, Aduszkiewicz:2015jna} for detailed description). The chosen coordinate system is drawn on the lower left: its origin lies in the middle of the VTPC-2, on the beam axis. 
  }
  \label{fig:detector-setup}
\end{figure*}

The \NASixtyOne experiment~\cite{Abgrall:2014xwa} uses a large acceptance hadron spectrometer located in the CERN North Area. The schematic layout of the \NASixtyOne detector is shown in Fig.~\ref{fig:detector-setup}. The detailed description of the full detector can be found in Ref.~\cite{Abgrall:2014xwa}. Here only the detector components, which were used in this analysis, are described.

A set of scintillation and Cherenkov counters as well as beam position detectors (BPDs) upstream of the spectrometer provide timing reference, identification and position measurements of incoming beam particles. The trigger scintillator counter S4 placed downstream of the target is used to select events with collisions in the target area by the absence of a charged particle hit.

Secondary beams of positively charged hadrons at 158~\GeVc are produced from 400~\GeVc protons extracted from the SPS accelerator. Particles of the secondary hadron beam are identified by two Cherenkov counters, a CEDAR~\cite{Bovet:1982xf} (for 158~\GeVc beam CEDAR-N) and a threshold counter (THC). The CEDAR counter, using a coincidence of six out of the eight photo-multipliers placed radially along the Cherenkov ring, provides positive identification of protons, while the THC, operated at pressure lower than the proton threshold, is used in anti-coincidence in the trigger logic. 
 A selection based on signals from the Cherenkov counters allowed one to identify beam protons with a purity of about 99\%. A consistent value for the purity was found by bending the beam into the TPCs with the full magnetic field and using identification based on its specific ionization energy loss \dedx~\cite{Claudia}.

The main tracking devices of the spectrometer are four large volume Time Projection Chambers (TPCs). Two of them, the \textit{vertex} TPCs (\mbox{VTPC-1} and \mbox{VTPC-2}), are located in the magnetic fields of two super-conducting dipole magnets with a combined bending power of 9~Tm which corresponds to about 1.5~T and 1.1~T fields in the upstream and downstream magnets, respectively. 

Two large \textit{main} TPCs (\mbox{MTPC-L} and \mbox{MTPC-R}) are positioned downstream of the magnets symmetrically to the beam line. The fifth small TPC (GAP TPC) is placed between \mbox{VTPC-1} and \mbox{VTPC-2} directly on the beam line. It closes the gap between the beam axis and the sensitive volumes of the other TPCs. The TPCs are filled with Ar and CO$_2$ gas mixtures. Particle identification in the TPCs is based on measurements of the specific energy loss (\dedx) in the chamber gas.

The p+p data sets, which are the topic of this paper, were recorded with the proton beam incident on a liquid hydrogen target (LHT), a 20~cm long cylinder positioned about 80~cm upstream of \mbox{VTPC-1}.

%%% Local Variables: 
%%% mode: latex
%%% TeX-master: "main"
%%% End: 

\section{Data sets and analysis technique}\label{sec:method}

\subsection{Data sets}
\begin{figure*}
  \centering
\includegraphics[width=0.49\textwidth]{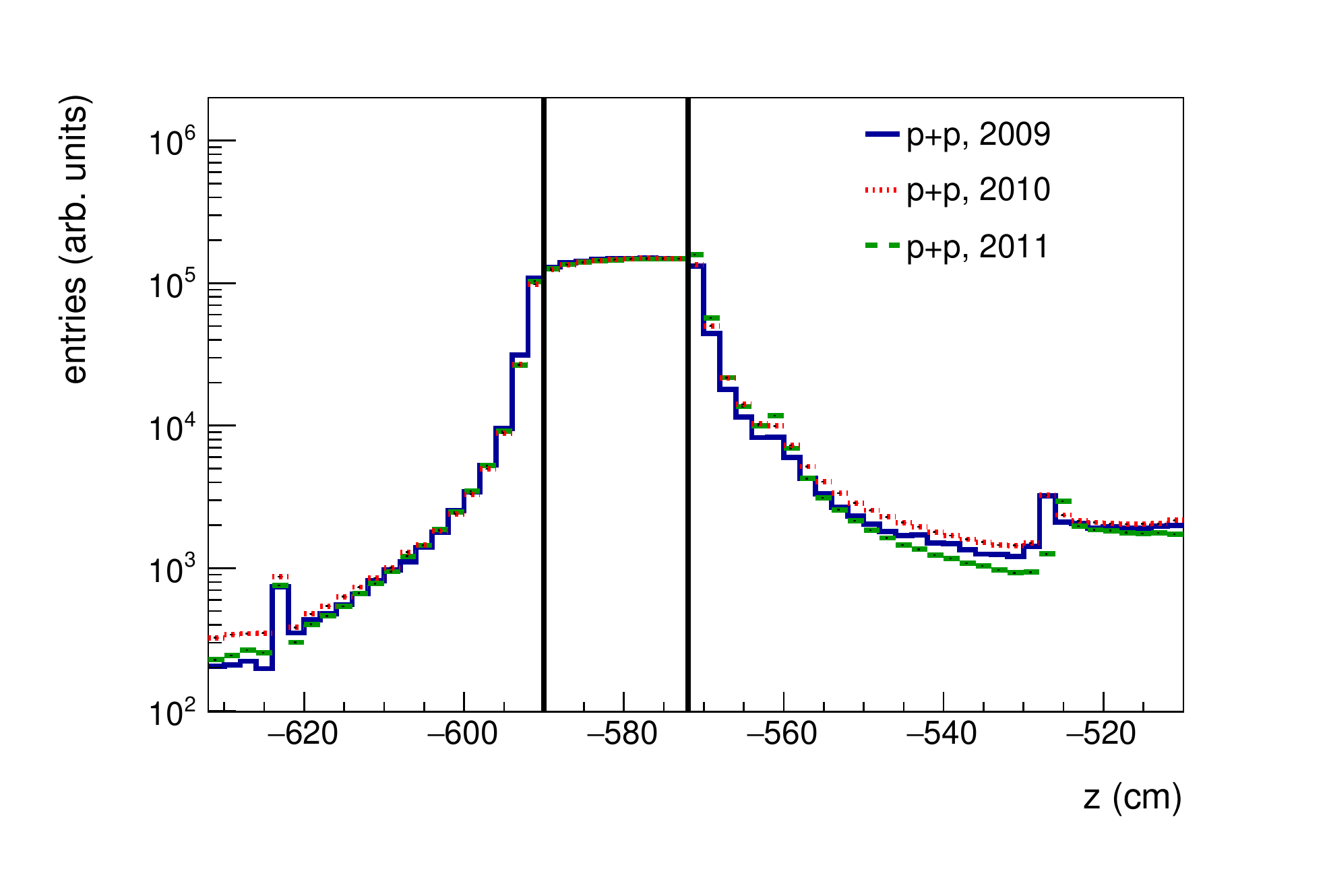}
\includegraphics[width=0.49\textwidth]{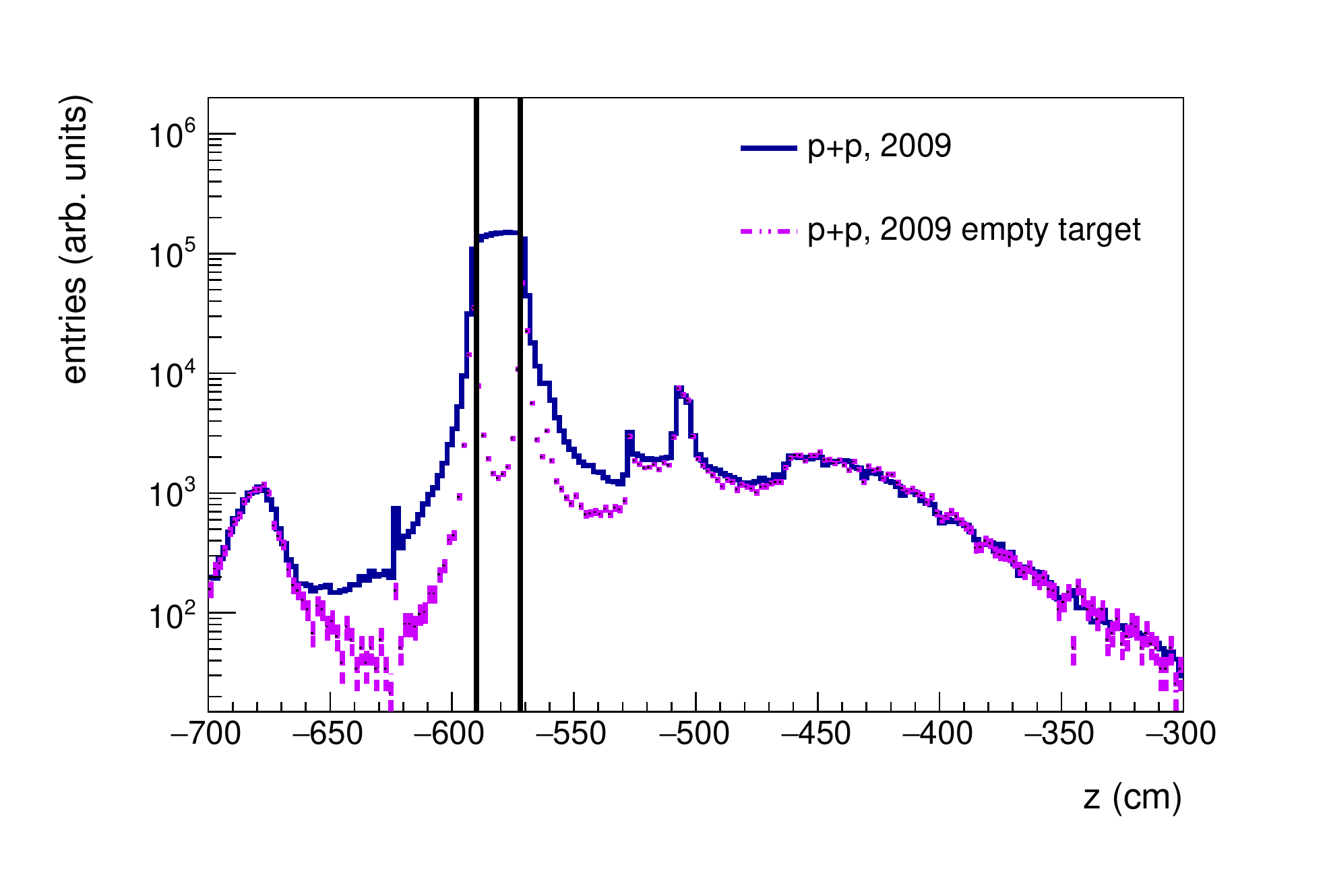}
\vspace{-0.5cm}
  \caption[]{(Color online) Left: Distributions of the $z$-coordinate of the reconstructed interaction vertex ($z$) for events recorded with the target inserted (2009, 2010 and 2011 data). Histograms are normalized to the same integral in the range $-590 < z < -572$ cm. Right: Distributions of the $z$-coordinate of the reconstructed interaction vertex for target-inserted (solid histogram) and target-removed (dash-dotted histogram) 2009 data. Histograms are normalized to the same integral in the range $-450 < z < -300$ cm. All event cuts were applied (see Sec.~\ref{s:event_selection}) with exception of cut (ii) and (v). Black vertical lines indicate the cuts used for the analysis (see Sec.~\ref{s:event_selection}). }
  \label{fig:vertexz}
\end{figure*}

The results for p+p interactions are based on high-statistics data runs (in years 2009, 2010, and 2011) which recorded about $56.65 \times 10^6$ collisions ($52.53 \times 10^6$ events selected by the interaction trigger)
of the proton beam with a 20~\cm long liquid hydrogen target (LHT). The conditions during the three runs were very similar as demonstrated in Fig.~\ref{fig:vertexz} (left) where the $z$-position (along the beam line) of the reconstructed p+p interaction vertex is shown. For the analysis the range of $z$-position of the main vertex was selected to cover mostly the LHT (see Sec.~\ref{s:event_selection}) in order to maximize the number of good events and minimize the contamination by off-target interactions. Figure~\ref{fig:vertexz} (right) shows that for the 2009 production the ratio of the number of events in the target-removed sample to the number of events in the target-inserted sample (ratio calculated in the range $-590 < z < -572$ cm; histograms normalized in the range $-450 < z < -300$ cm) is on the level of 4.8\%, and therefore no correction for non-target interactions was applied.
An alternative method of analysis (see for example Ref.~\cite{Abgrall:2013qoa}) would be to measure and subtract the resonance yields in the target-removed data, but both the \textit{standard} method and the \textit{template}-fitting method used in this paper cannot be applied to data sets with small statistics such as the target removed data recorded by \NASixtyOne.
In order to estimate the systematic biases related to the contamination by off-target interactions the window of $z$-position of the main vertex was varied (see Sec.~\ref{s:statistical_and_systematic_uncertainties}).

Table~\ref{tab:data_sets} presents the details of data sets collected in the three separate data taking periods. The number of events recorded with the interaction trigger, as well as the number of events selected for the analysis (see Sec.~\ref{s:event_selection}) are shown. One sees that only 44--56\% of the events were used for the analysis. This drop is caused mainly by BPD reconstruction inefficiencies and off-target interactions accepted by the trigger.  
The number of tracks, given in the Table~\ref{tab:data_sets}, refers to tracks registered in accepted events only. The agreement of the fractions of accepted tracks in the three analyzed data sets confirms the similarity of the data recorded in 2009, 2010 and 2011. For the analysis of $K^*(892)^0$ production these three data sets were combined at the level of preparing invariant mass distributions (Sec.~\ref{s:signal_extraction}).

\begin{table} [h]
\small
	\centering
	\begin{tabular}{|c|c|c|c|c|}
		\hline
		& 2009 & 2010 & 2011 & Total \\
		\hline
		Number of events & 2.87M (100\%) & 37.78M (100\%) & 11.88M (100\%) & 52.53M (100\%) \\
		selected by interaction trigger & & & & \\
		\hline
		Number of events after cuts & 1.26M (43.9\%) & 19.97M (52.9\%) & 6.62M (55.7\%) & 27.85M (53.0\%)\\
		\hline
		Number of tracks & 8.62M (100\%) & 136.58M (100\%) & 45.48M (100\%) & 190.68M (100\%) \\
		\hline
		Number of tracks after cuts & 4.81M (55.8\%) & 76.41M (55.9\%)& 24.91M (54.8\%) & 106.13M (55.7\%)\\
		without \dedx cut & & & & \\
		\hline
		Number of tracks after all cuts & 2.26M (26.2\%) & 35.79M (26.2\%)& 11.74M (25.8\%) & 49.79M (26.1\%)\\
		\hline
	\end{tabular}
\caption{Data sets used for the analysis of $K^*(892)^0$ production. The same event and track cuts (Sec.~\ref{s:event_selection}, \ref{s:track_selection} and \ref{s:dedx_cuts}) were used for all three data taking periods. }
\label{tab:data_sets}
\end{table}

\subsection{Analysis method}
\label{sec:analysis_method}

The details of \NASixtyOne calibration, track and vertex reconstruction procedures, as well as simulations used to correct the reconstructed data, are discussed in Refs.~\cite{Abgrall:2013qoa, Aduszkiewicz:2015jna, Aduszkiewicz:2016mww}. 
In the following section the analysis technique developed for the measurement of the $K^*(892)^0$ spectra in p+p interactions is described. The procedure used for the data analysis consists of the following steps:

\begin{itemize}
	\item [(i)] application of event and track selection criteria,
	\item [(ii)] selection of $K^{+}$ and $\pi^{-}$ candidates based on the measurement of their ionization energy loss (\dedx) in the gas volume of the TPCs,
	\item [(iii)] creation of invariant mass distribution of $K^{+} \pi^{-}$ pairs,
	\item [(iv)] creation of invariant mass distribution of $K^{+} \pi^{-}$ pairs for mixed events and Monte Carlo templates,
	\item [(v)] extraction of $K^*(892)^0$ signal, 
	\item [(vi)] application of corrections (obtained from simulations) to the raw numbers of $K^*(892)^0$; they include losses of inelastic p+p interactions due to the on-line and off-line event selection as well as losses of $K^*(892)^0$ due to track and pair selection cuts and the detector geometrical acceptance.
\end{itemize}

The details of the steps are described in the following subsections.

%%%%%%%%%%%%%%%%%%%%%%%%%%%%%%%%%%%%%%%%%%%%%%%%%%%%%%%
\subsection{Event selection}
\label{s:event_selection}

Inelastic p+p interactions were selected by the following criteria:

\begin{itemize}
	\item [(i)] an interaction was recognized by the trigger logic (see Refs.~\cite{Abgrall:2013qoa, Aduszkiewicz:2015jna} for detailed description),
	\item [(ii)] no off-time beam particle was detected within $\pm 1$ $\mu$\s around the trigger (beam) particle,
	\item [(iii)] the trajectory of the beam particle was measured in at least one of BPD-1 or BPD-2 and in the BPD-3 detector and was well reconstructed,
	\item [(iv)] the primary interaction vertex fit converged,
	\item [(v)] the fit of the $z$-coordinate of the primary p+p interaction vertex (see Fig.~\ref{fig:vertexz}) converged and the fitted $z$ position was found between -590~\cm and -572~\cm, 
where the center of the LHT was at -580~\cm. The range of this cut was selected to maximize the number of good events and minimize the contamination by off-target interactions, 
	\item [(vi)] events with a single, well measured positively charged track with absolute momentum close to the beam momentum ($p > p_{beam}-1$ \GeVc) were rejected.   
\end{itemize} 

The above event cuts select well measured inelastic p+p interactions. 
The background due to elastic interactions is removed (cuts (iv) and (vi)).
The contribution of off-target interactions is reduced (cut (v)). The losses of inelastic interactions due to the event selection procedure were corrected using simulations (see below).
The number of events after these cuts is $27.85 \times 10^6$.

\subsection{Track selection}
\label{s:track_selection}

After the event selection criteria a set of track quality cuts were applied to individual tracks. These cuts were used to ensure high reconstruction efficiency, proper identification of tracks and to reduce the contamination of tracks from secondary interactions, weak decays and off-time interactions.   
The individual tracks were selected by the following criteria:

\begin{itemize}
	\item [(i)] the track fit including the interaction vertex converged,
	\item [(ii)] the total number of reconstructed points on the track should be greater than 30, 
	\item [(iii)] the sum of the number of reconstructed points in VTPC-1 and VTPC-2 was greater than 15 or the number of reconstructed points in the GAP TPC was greater than 4,
	\item [(iv)] the distance between the track extrapolated to the interaction plane and the interaction point (impact parameter) should be smaller than 4~\cm in the horizontal (bending) plane and 2~\cm in the vertical (drift) plane,
	\item [(v)] the track momentum (in the laboratory reference system) is in the range $3 \leq p_{lab} \leq 158$~\GeVc, 
	\item [(vi)] the track transverse momentum is required to be smaller than 1.5~\GeVc,
	\item [(vii)] \dedx track cuts were applied to select $K^{+}$ and $\pi^{-}$ candidates (see Sec.~\ref{s:dedx_cuts}).
\end{itemize} 

The number of tracks left after these cuts is about $49.79 \times 10^6$.

\subsection{Selection of kaon and pion candidates}
\label{s:dedx_cuts}

\begin{figure*}
  \centering
  \includegraphics[width=0.45\textwidth]{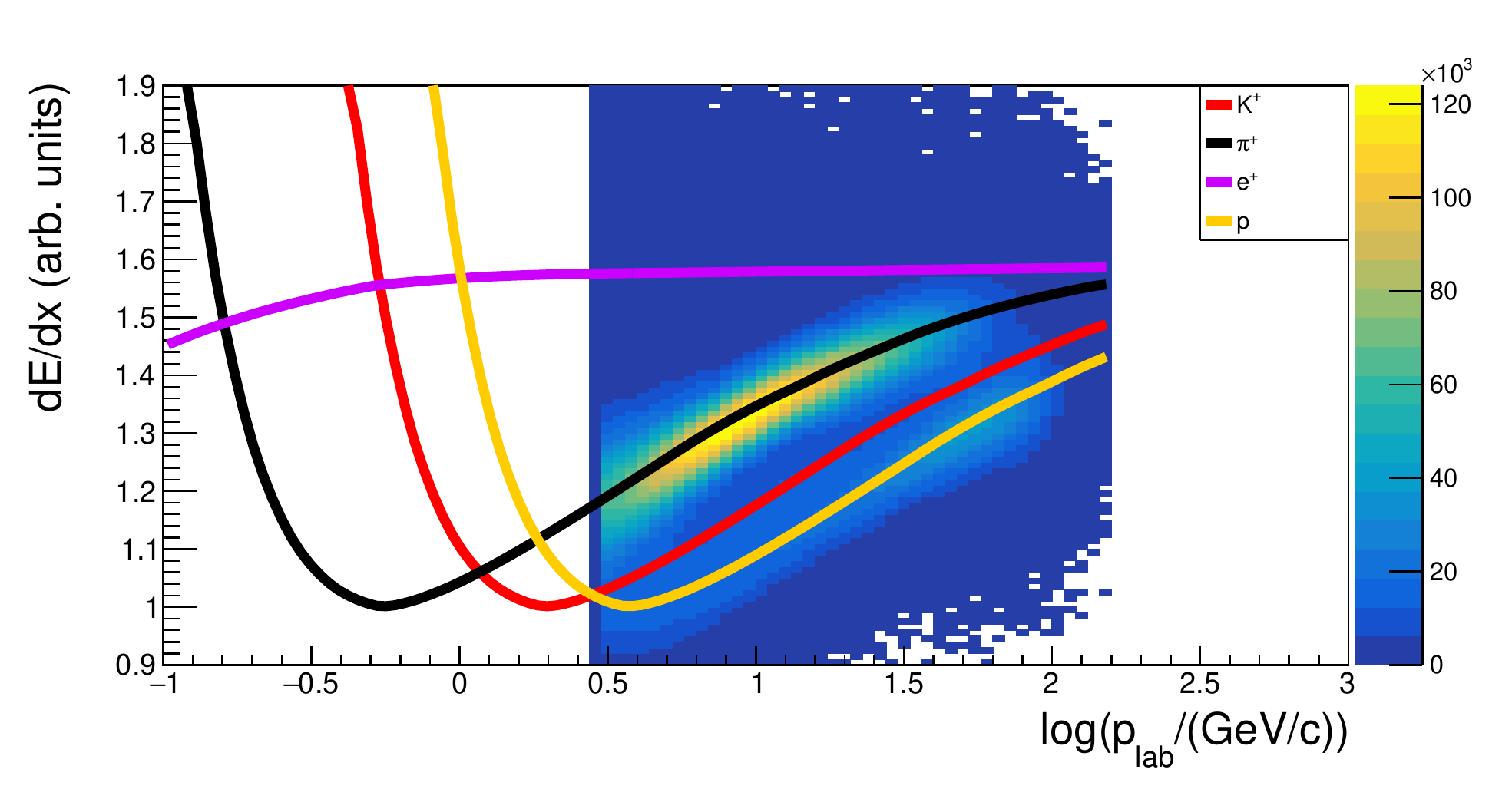}
 \includegraphics[width=0.45\textwidth]{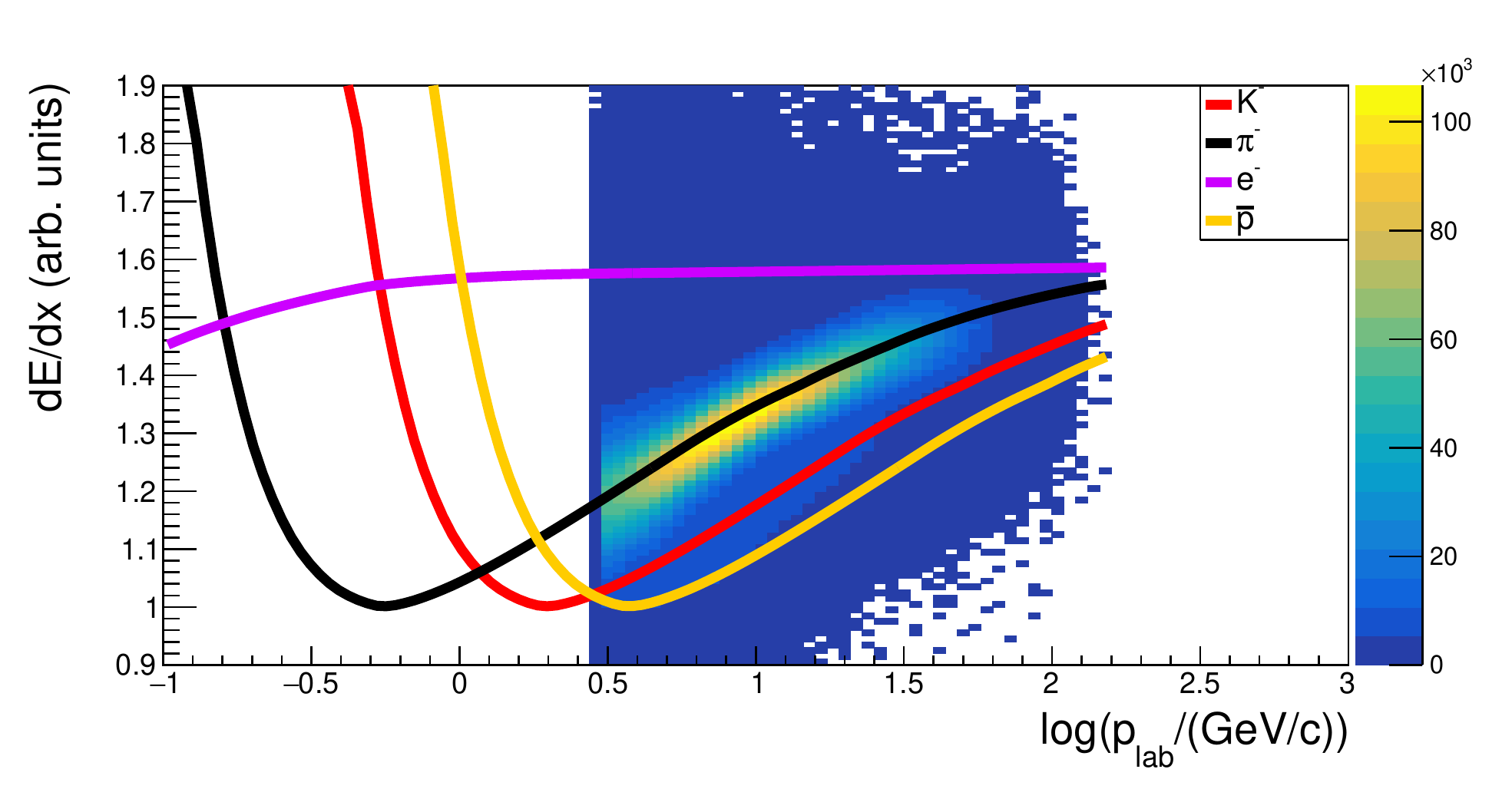} 
 \includegraphics[width=0.45\textwidth]{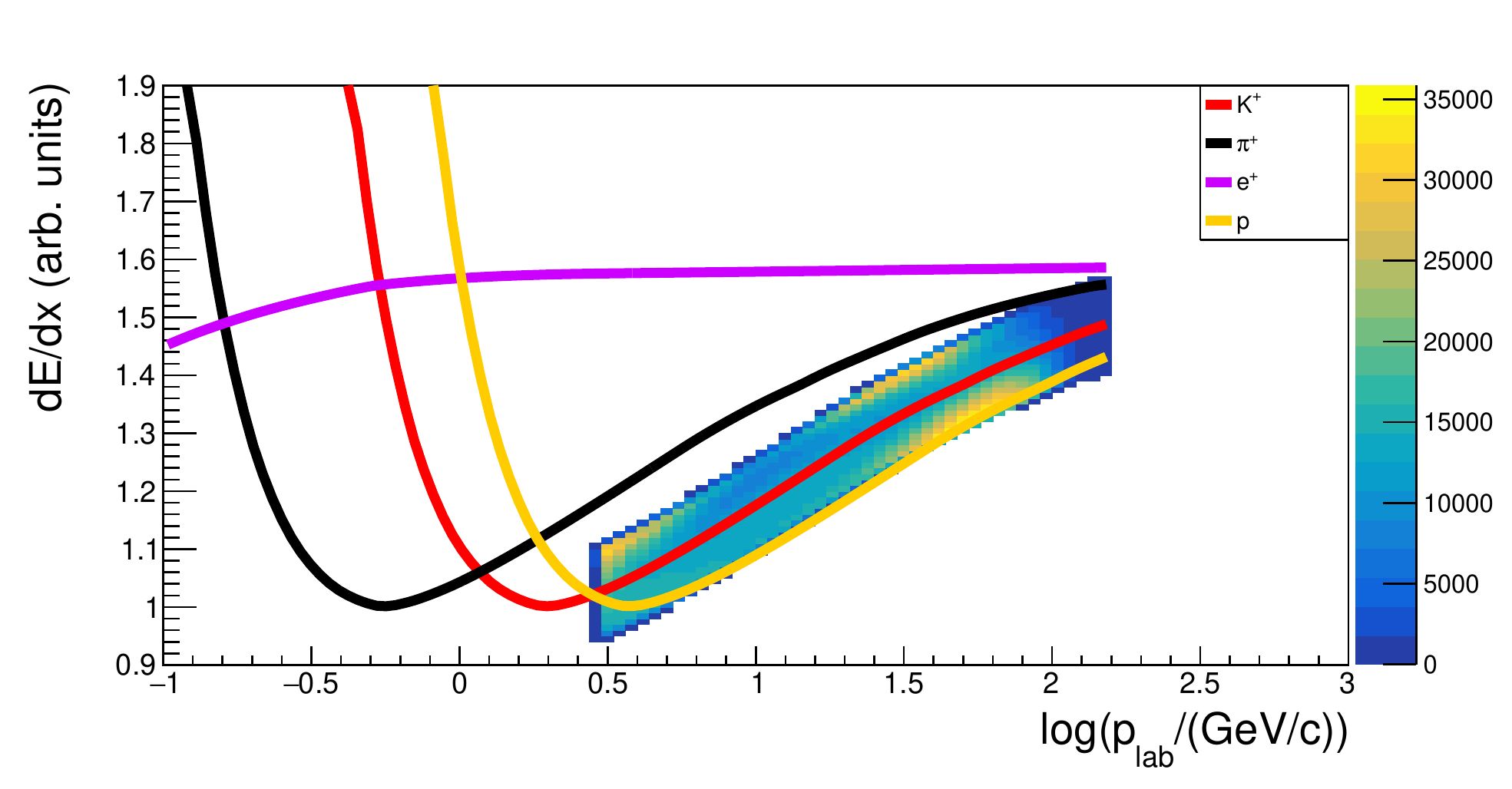}
  \includegraphics[width=0.45\textwidth]{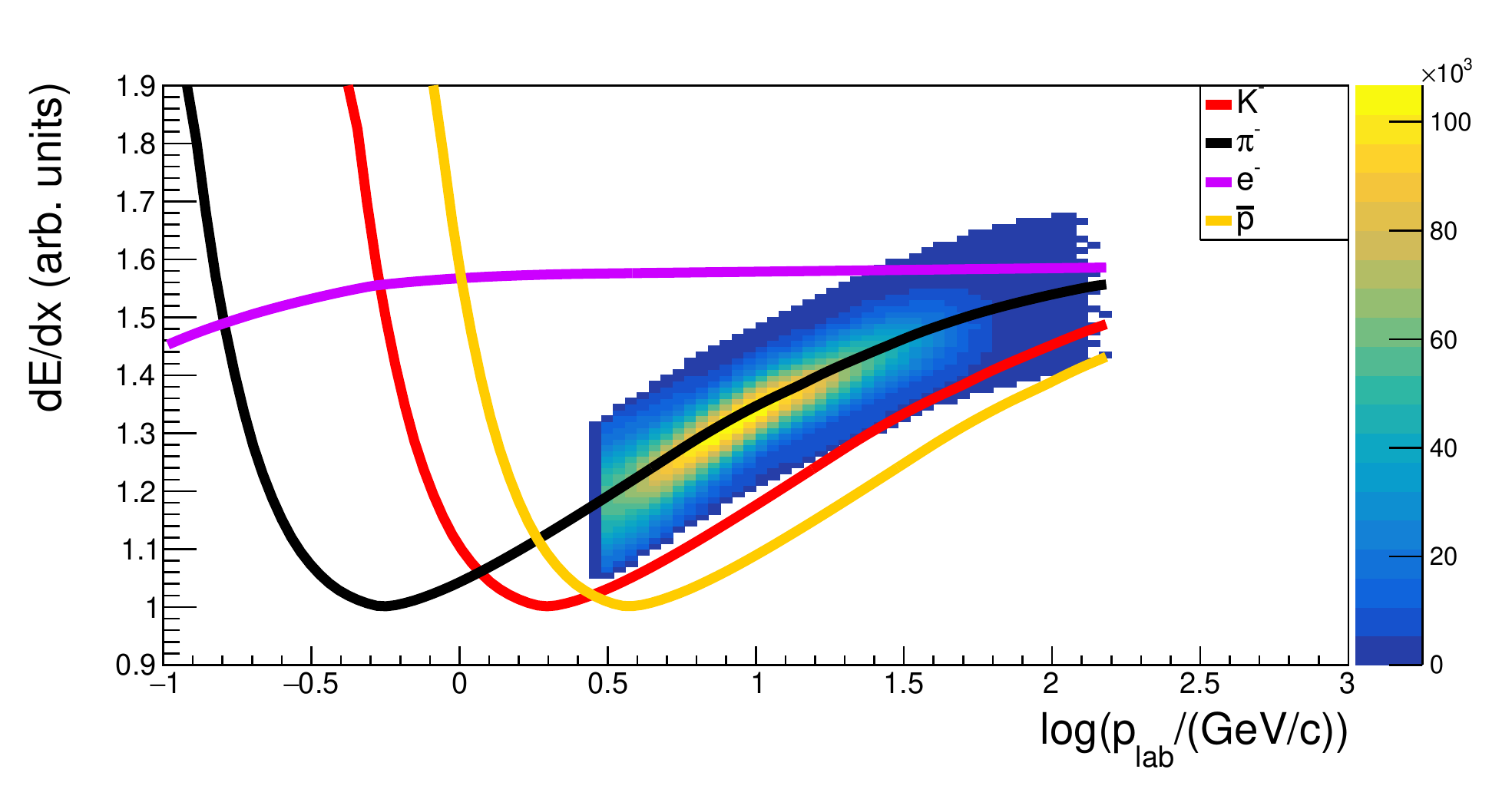}
  \caption[]{
    (Color online) Top: the values of \dedx versus $\log(p_{lab}/(\GeVc))$ for positively (left) and negatively (right) charged particles after track cuts (i) -- (vi) from Sec.~\ref{s:track_selection}. The Bethe-Bloch curves are also drawn. Bottom: selection of $K^{+}$ (left) and $\pi^{-}$ (right) candidates. }
  \label{fig:dEdx}
\end{figure*}

Charged particle identification in the \NASixtyOne experiment is based on the measurement of their ionization energy loss (\dedx) in the gas of the TPCs and of the time of flight (\tof) obtained from the ToF-L and ToF-R walls. For the region of the relativistic rise of the ionization at large momenta, the measurement of \dedx alone allows identification. At lower momenta the \dedx bands for different particle species overlap and the identification based only on measurements of \dedx in the TPCs (this analysis) is not enough. For this reason the track cut (v) was applied. In Fig.~\ref{fig:dEdx} the \dedx values as a function of total momentum ($p_{lab}$), measured in the laboratory reference system, are shown for positively and negatively charged particles, separately. The $K^{+}$ and $\pi^{-}$ candidates were selected by requiring their \dedx values to be within $1.5\sigma$ or $3.0\sigma$ around their nominal Bethe-Bloch values, respectively. Here $\sigma$ represents the typical standard deviation of a Gaussian fitted to the \dedx distribution of kaons and pions. Since only small variations of $\sigma$ were observed for different total momentum and transverse momentum bins, fixed values  $\sigma=0.044$ were used for $K^{+}$ and $\sigma=0.052$ for $\pi^{-}$. The bands of selected $K^{+}$ and $\pi^{-}$ candidates are shown in the bottom panel of Fig.~\ref{fig:dEdx}

\subsection{$K^*(892)^0$ signal extraction}
\label{s:signal_extraction}

The raw numbers of $K^{*}(892)^0$ are usually obtained by performing fits to the invariant mass spectra with the sum of a background and a signal function. The invariant mass is defined as:
\begin{equation}
m_{K^{+} \pi^{-}}=\sqrt{(E_{K^{+}}+E_{\pi^{-}})^2 - (\overrightarrow{p_{K^{+}}}+\overrightarrow{p_{\pi^{-}}})^2},
\end{equation}
where $E$ represents the total energy and $\vec{p}$ the momentum vector of daughter particles from $K^{*}(892)^0$ decay.

\begin{figure*}
  \centering
  \includegraphics[width=0.45\textwidth]{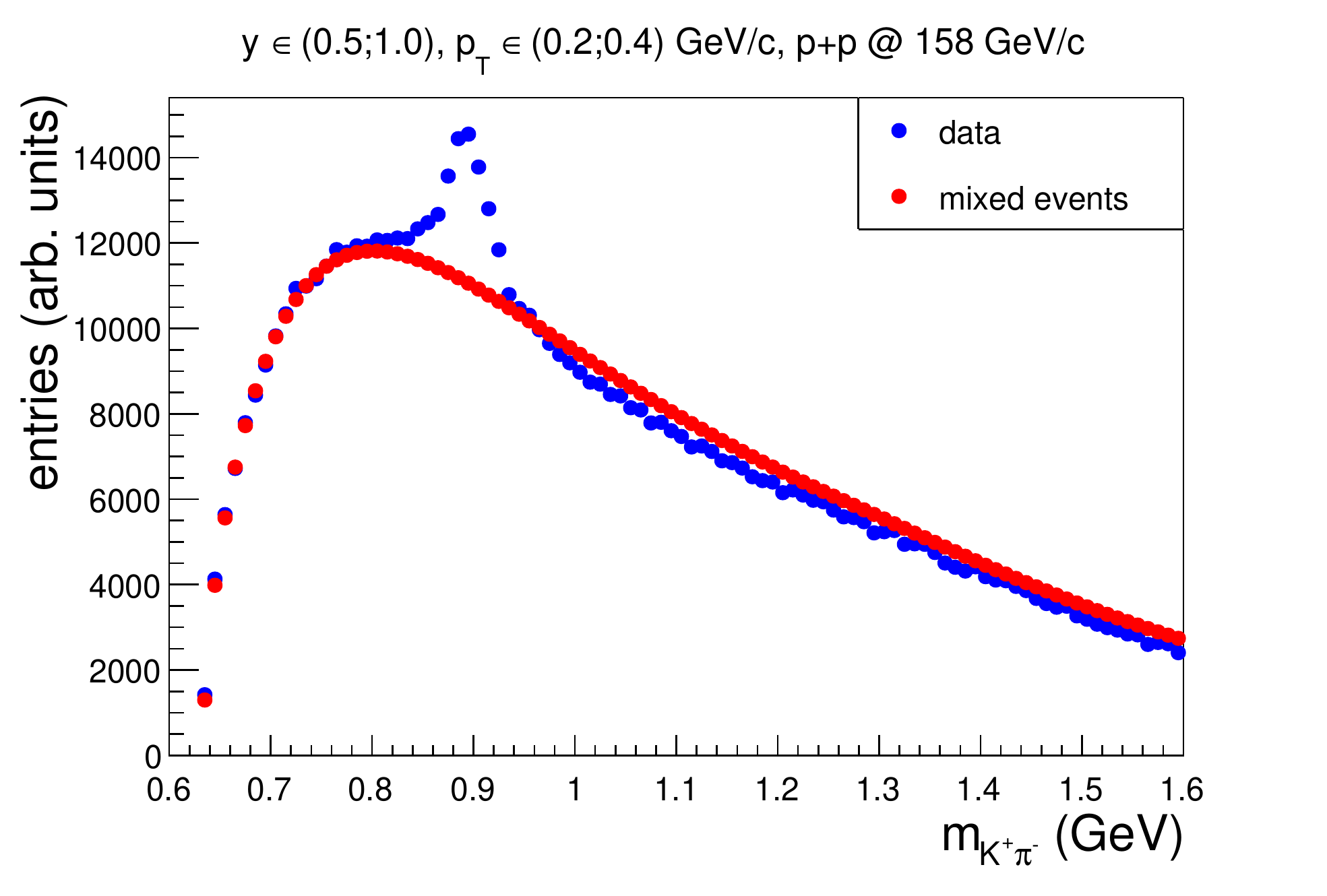}
 \includegraphics[width=0.45\textwidth]{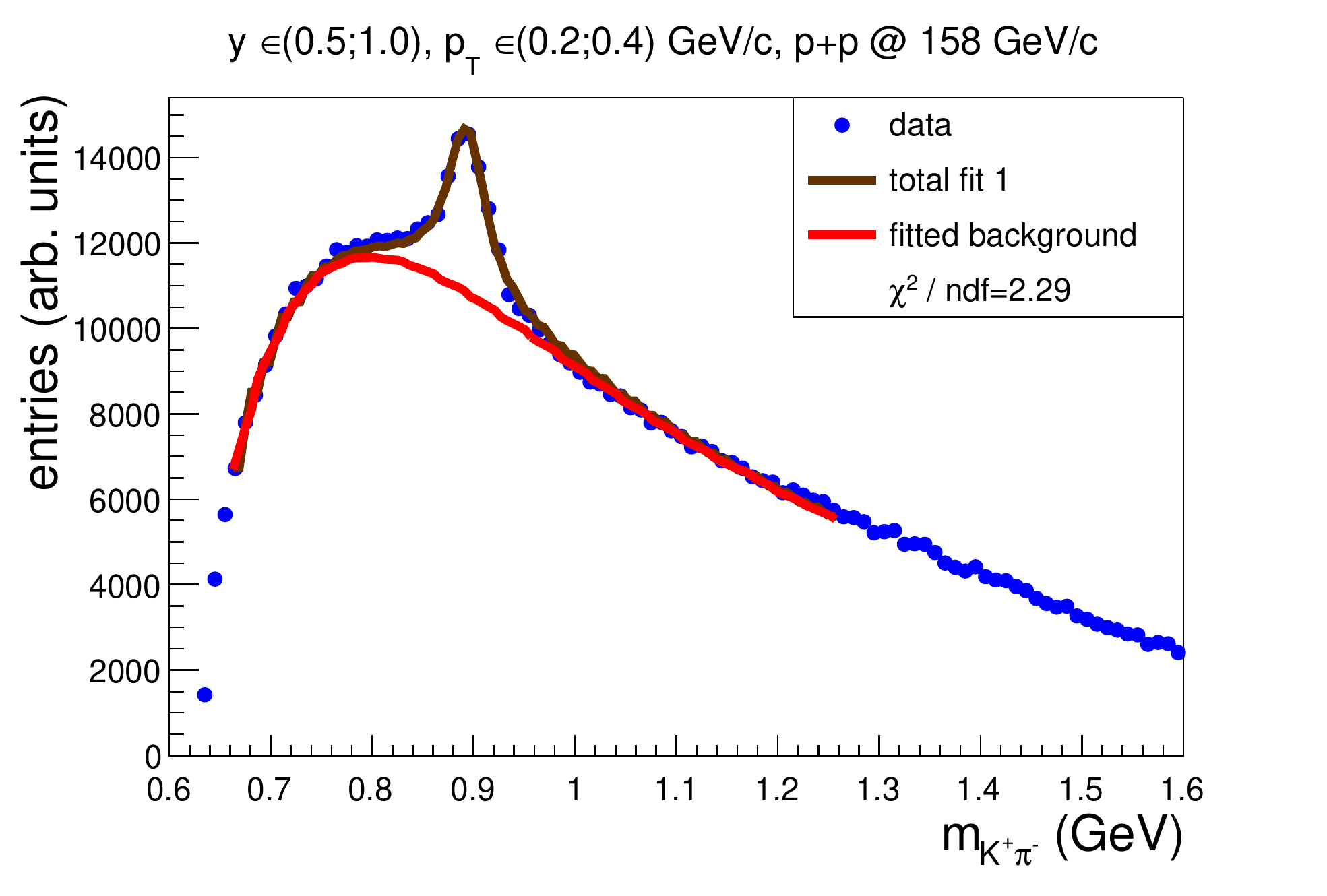} 
 \includegraphics[width=0.45\textwidth]{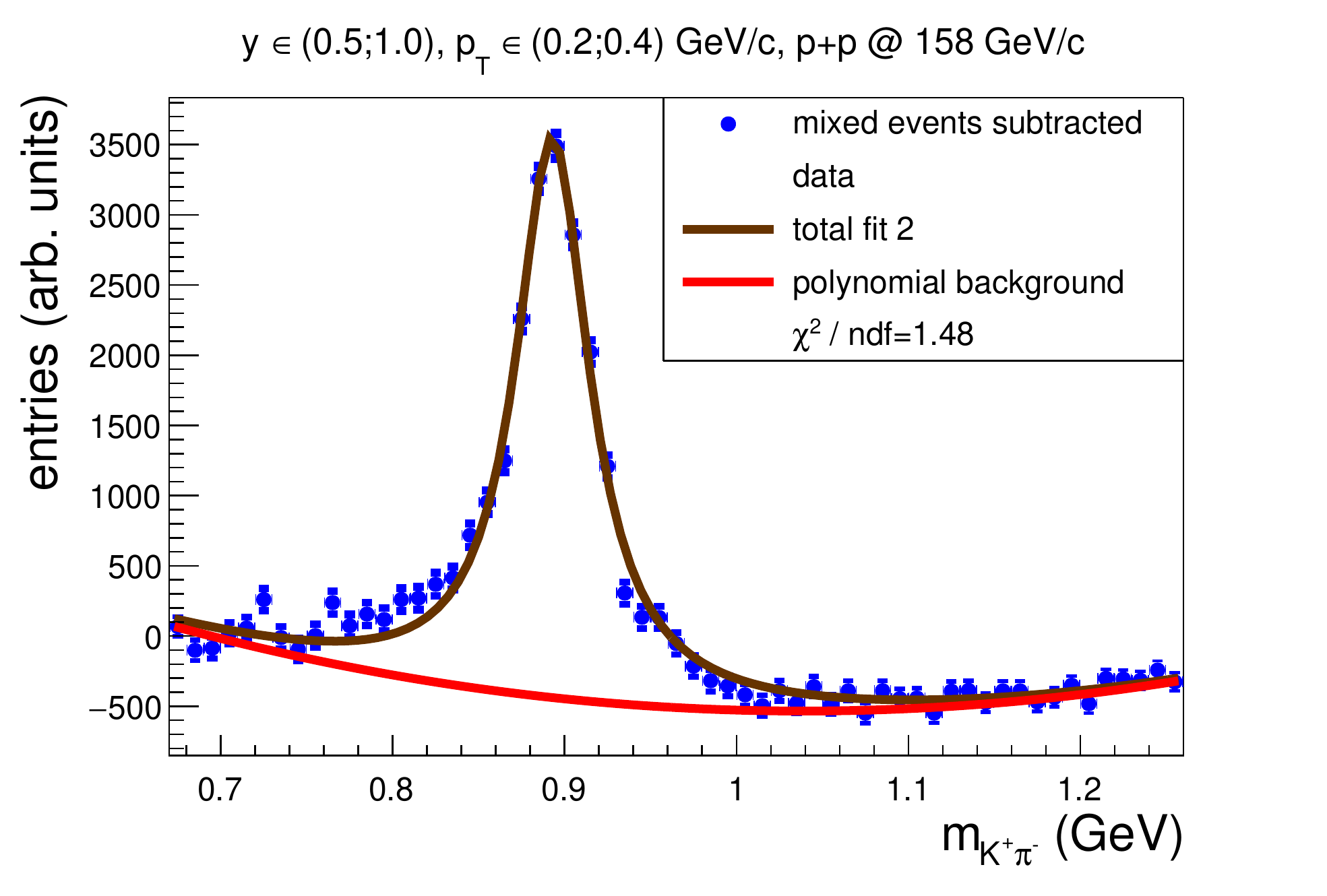}
  \includegraphics[width=0.45\textwidth]{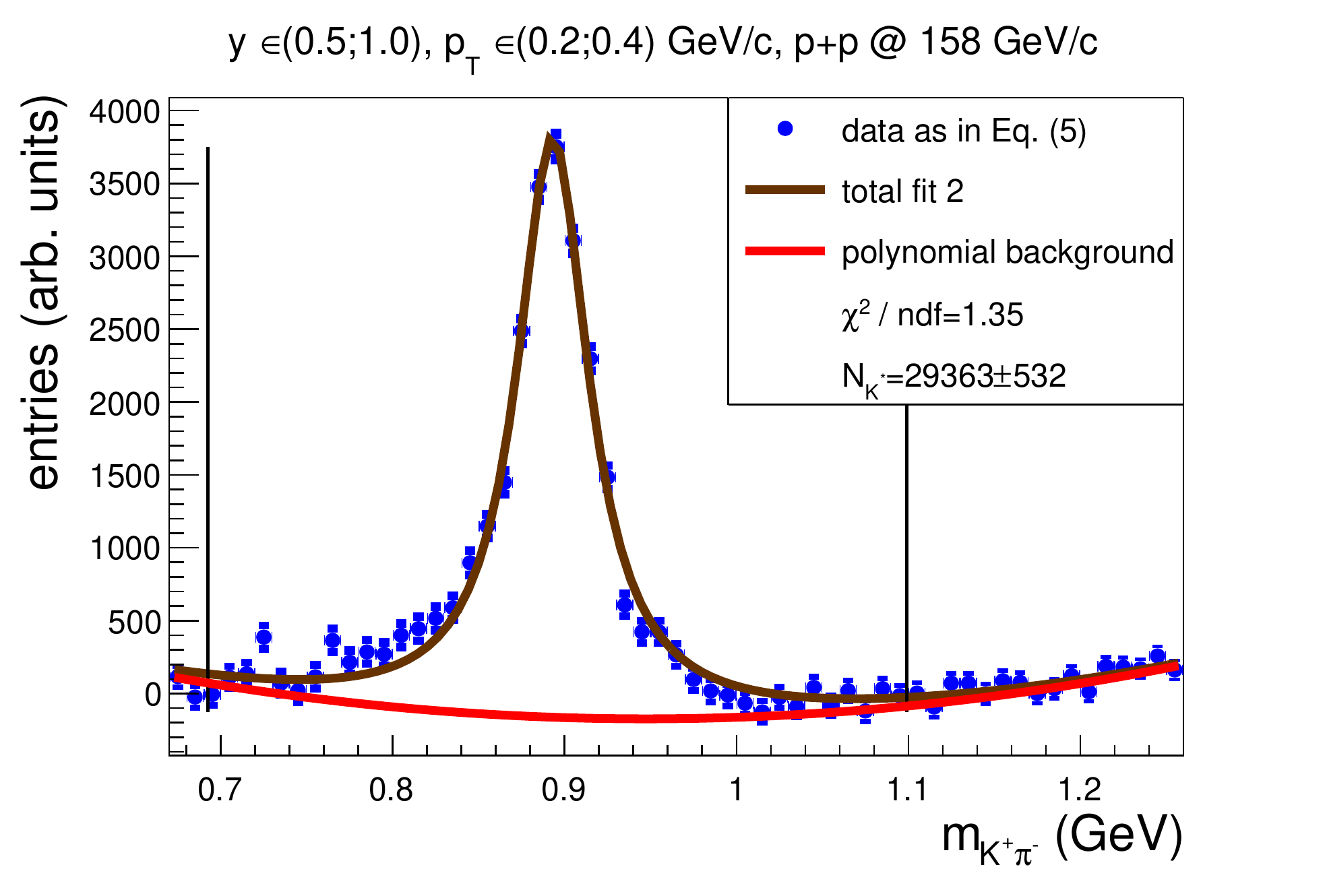}
  \caption[]{
    (Color online) The example of the procedure of signal extraction for $K^{*}(892)^0$ in rapidity bin $0.5<y<1.0$ (all rapidity values in the paper are given in the center-of-mass reference system) and transverse momentum bin $0.2<p_T<0.4$~\GeVc for p+p collisions at 158~\GeVc. Top, left: data signal (blue points), and background histogram (red points) obtained from mixed events (standard method). Top, right: data signal (blue points), and fitted background (red line) obtained from the templates. Bottom: background subtracted signal for the standard method (left) and template method (right) -- more details in the text. Thin black vertical lines in bottom right panel correspond to the range of integrating fit functions while obtaining the raw number of $K^{*}(892)^0$ mesons ($m_0 \pm 4\Gamma_0$; see the text for details). 
}
  \label{fig:SM_TM_comp}
\end{figure*}

\begin{figure*}
  \centering
  \includegraphics[width=0.45\textwidth]{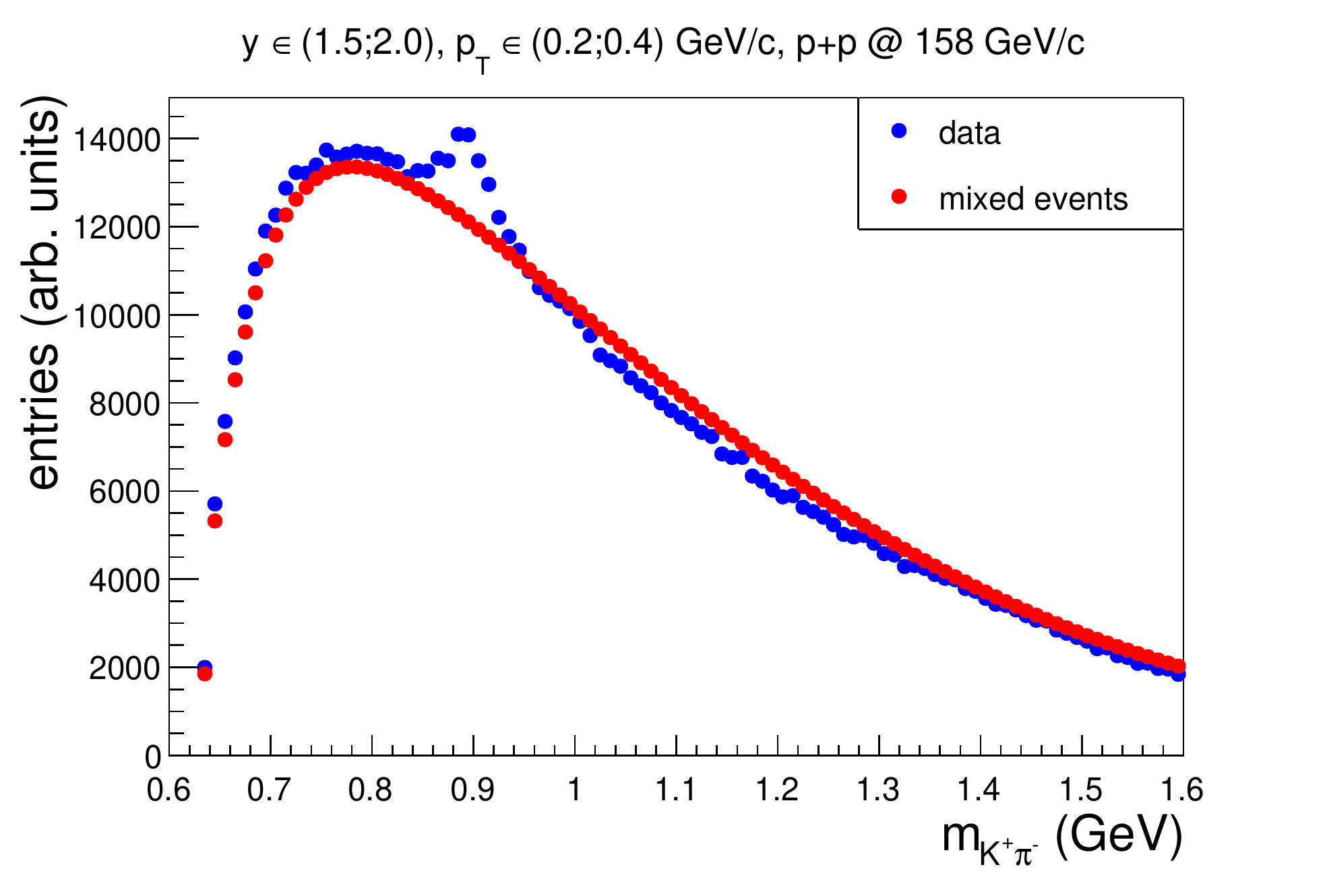}
 \includegraphics[width=0.45\textwidth]{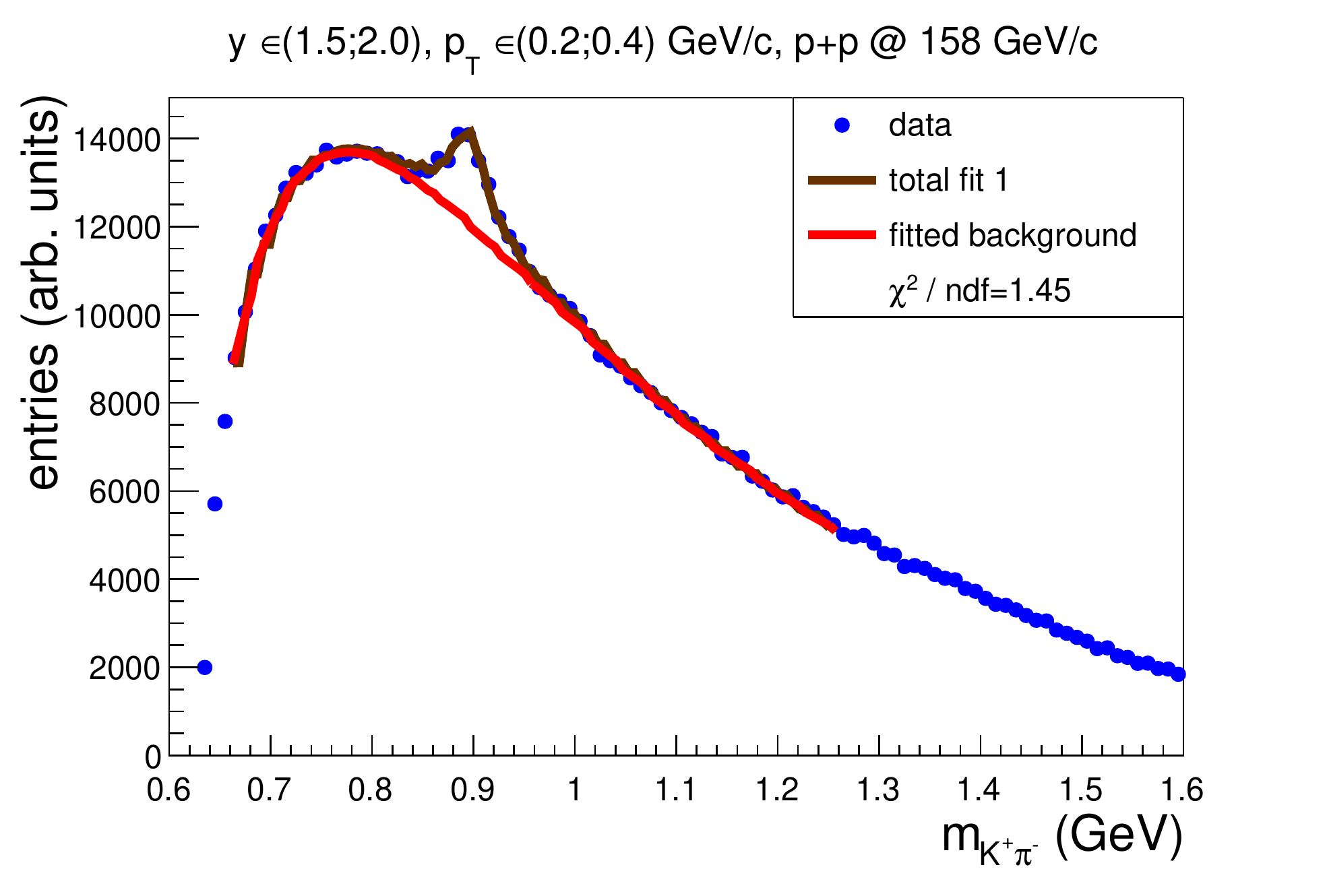} 
 \includegraphics[width=0.45\textwidth]{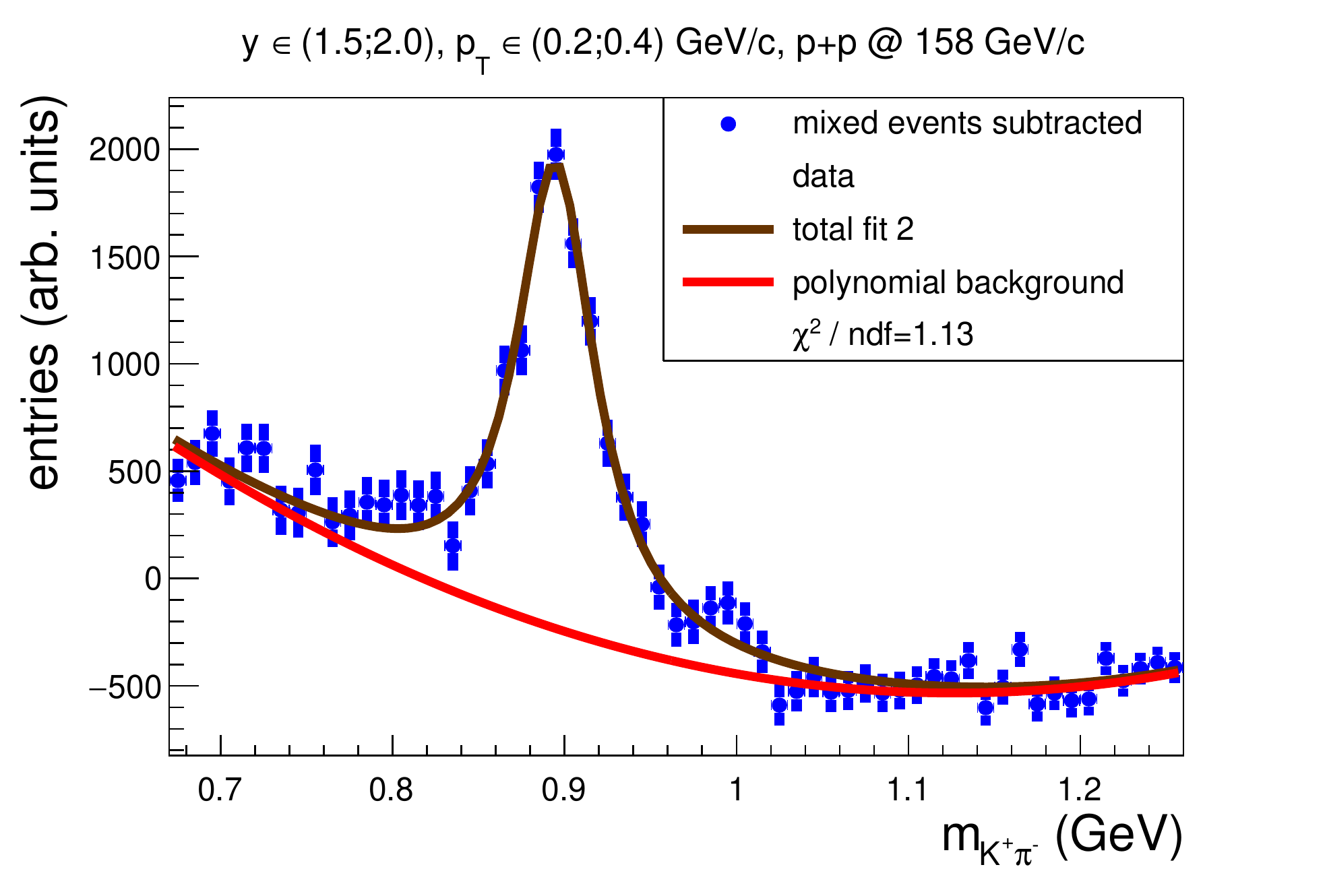}
  \includegraphics[width=0.45\textwidth]{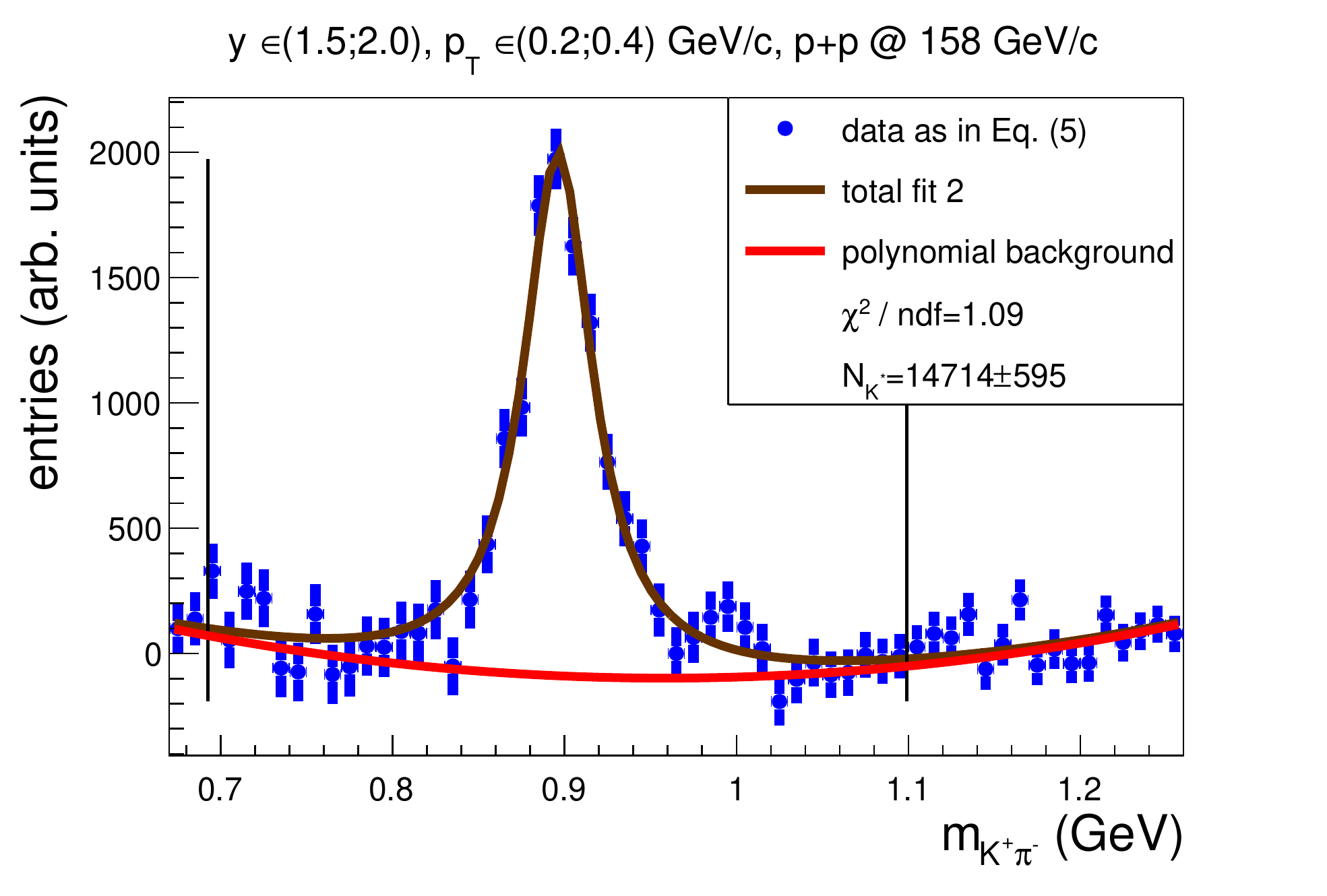}
  \caption[]{
    (Color online) Same as Fig.~\ref{fig:SM_TM_comp} but for $1.5<y<2.0$ and $0.2<p_T<0.4$~\GeVc. }
  \label{fig:SM_TM_comp_25}
\end{figure*}

In the \textit{standard method} (\textit{mixing method}) the large combinatorial background is estimated by invariant mass spectra calculated for $K^{+} \pi^{-}$ pairs originating from different events. Figures~\ref{fig:SM_TM_comp} and \ref{fig:SM_TM_comp_25} (top, left) show combinatorial background histograms (red points) compared to the data histograms of $m_{K^{+} \pi^{-}}$ (blue points). Mixed events were normalized to the same number of pairs as in real data in the invariant mass range from 0.6 to 1.6~\GeV.   
After subracting the normalized mixed event background the blue points in Figs.~\ref{fig:SM_TM_comp}, \ref{fig:SM_TM_comp_25} (bottom, left) were obtained. The $K^{*}(892)^0$ signal is prominently seen, but the histogram still shows a residual background, seen especially for low invariant mass values. This residual background probably comes from the products of other resonance decays, which are not properly accounted for by the event-mixing, and should be subtracted. The final fit (\textit{total fit 2}) was performed with the function of Eq.~(\ref{eq:total_fit_2}) using an additional background component based on a second order polynomial: 
 
\begin{equation}
f (m_{K^{+}\pi^{-}}) = d \cdot {(m_{K^{+}\pi^{-}})}^2 + e \cdot (m_{K^{+}\pi^{-}}) + f + g \cdot BW (m_{K^{+}\pi^{-}}),
\label{eq:total_fit_2}
\end{equation}   
where $d$, $e$, $f$, and $g$ are free parameters of the fit, and the Breit-Wigner ($BW$) component is described by Eq. (\ref{eq:BW_function}):
\begin{equation}
BW(m_{K^{+}\pi^{-}}) = A \cdot \frac{\frac{1}{4} \cdot \Gamma_{K^*}^2}{(m_{K^{+}\pi^{-}} - m_{K^*})^2 + \frac{1}{4} \Gamma_{K^*}^2},
\label{eq:BW_function}
\end{equation}
where $A$ is the normalization factor, and $m_{K^*}$ and $\Gamma_{K^*}$ are also fitted. The initial values of the mass ($m_{K^*}$) and width ($\Gamma_{K^*}$) parameters of $K^{*}(892)^0$ were taken from the Particle Data Group (PDG): $m_{K^*}=m_0=0.89555$~\GeV and $\Gamma_{K^*}=\Gamma_0=0.0473$~\GeV~\cite{PDG}. 
 The red lines (\textit{polynomial background}) in Figs.~\ref{fig:SM_TM_comp}, \ref{fig:SM_TM_comp_25} (bottom, left) show the fitted additional background component (Eq.~(\ref{eq:total_fit_2}) without $BW$) and the brown lines (\textit{total fit 2}) the total fit result (Eq.~(\ref{eq:total_fit_2})).

In order to obtain a better background description compared to the mixing method, the \textit{template method} was applied. Namely, the invariant mass spectra of the data (blue data points in Figs.~\ref{fig:SM_TM_comp}, \ref{fig:SM_TM_comp_25} (top, right)) were fitted with a function given by Eq.~(\ref{eq:minv_function}):

\begin{equation}
f (m_{K^{+}\pi^{-}}) = a \cdot T_{res}^{MC} (m_{K^{+}\pi^{-}}) + b \cdot T_{mix}^{DATA} (m_{K^{+}\pi^{-}}) + c \cdot BW (m_{K^{+}\pi^{-}}).
\label{eq:minv_function}
\end{equation}   

The background is described as a sum of two contributions: $T_{res}^{MC}$ and $T_{mix}^{DATA}$. $T_{mix}^{DATA}$ is the background estimated based on the mixing method, which was discussed above. 
The $T_{res}^{MC}$ template (MC stands for Monte Carlo) is the shape of background, which describes the contribution of $K^{+} \pi^{-}$ pairs originating from:

\begin{itemize}
	\item [(i)] combination of tracks that come from decays of resonances different than $K^{*}(892)^0$, for example one track from a $\rho^0$ meson and one from a $K^{*+}$ meson,
	\item [(ii)] combination of tracks where one comes from decay of a resonance and one comes from direct production in the primary interaction.
\end{itemize}

The $T_{res}^{MC}$ templates were constructed by passing p+p interactions, generated with the \EposLong~\cite{Werner:2005jf} hadronic interaction model using the CRMC 1.4 package~\cite{EPOS_CRMC}, through the \NASixtyOne detector simulation chain and then through the same reconstruction routines as the data. The simulation keeps the history of particle production thus allowing to identify their identity and origin enabling the construction the proper templates. 
For the reconstructed MC events, the same event and track selection criteria, as for real data, were used. They also include the effects of the limited acceptance of the detector. Both the template and the data histograms were computed in bins of rapidity $y$ (calculated in the center-of-mass reference system) and transverse momentum $p_T$.  

Finally, the signal ($BW$) is described using the Breit-Wigner distribution Eq.~(\ref{eq:BW_function}).

The $T_{res}^{MC}$ and $T_{mix}^{DATA}$ histograms in the fit function Eq.~(\ref{eq:minv_function}) were normalized to have the same numbers of pairs as the real data histogram in the invariant mass range from 0.6 to 1.6~\GeV. 
The symbols $a$, $b$ and $c$ in Eq.~(\ref{eq:minv_function}) are the normalization parameters of the fit ($a+b+c=1$), which describe the contributions of $T_{res}^{MC}$, $T_{mix}^{DATA}$ and $BW$ to the invariant mass spectra. The mass and width of the $K^{*}(892)^0$ are the parameters of the Breit-Wigner shape obtained within the mass window $m_0 \pm 4\Gamma_0$. The values from \textit{total fit 2} (see Fig.~\ref{fig:SM_TM_comp} or \ref{fig:SM_TM_comp_25} (bottom, right)) were used in the results section below.

In Figs.~\ref{fig:SM_TM_comp}, \ref{fig:SM_TM_comp_25} (top, right), the fitted invariant mass spectra, using Eq.~(\ref{eq:minv_function}), are presented by brown curves (\textit{total fit 1}). The red lines (\textit{fitted background}) show the fitted function without the signal contribution ($BW$). The fits (brown and red curves) were performed in the invariant mass range from 0.66~\GeV to 1.26~\GeV. It is seen that Eq.~(\ref{eq:minv_function}) (without $BW$ component) describes the background much better than only mixed events (Figs.~\ref{fig:SM_TM_comp}, \ref{fig:SM_TM_comp_25} (top, left)). After MC template and mixed event background subtraction (see Eq. (\ref{eq:Nbin})),
the resulting mass distributions (blue data points) are shown in Figs.~\ref{fig:SM_TM_comp}, \ref{fig:SM_TM_comp_25} (bottom, right). One sees that the remaining background (red curves) is much less significant than in the case of the standard method (Figs.~\ref{fig:SM_TM_comp}, \ref{fig:SM_TM_comp_25} (bottom, left)). In fact, a small residual background is present mostly for the $y$ and $p_T$ bins in which the statistics is very low. To subtract it, a fit of the blue histograms was performed as the last step using the function Eq.~(\ref{eq:total_fit_2}). The results are shown in Figs.~\ref{fig:SM_TM_comp}, \ref{fig:SM_TM_comp_25} (bottom, right). The red lines (\textit{polynomial background}) illustrate the remaining residual background (Eq.~(\ref{eq:total_fit_2}) without $BW$) and the brown curves (\textit{total fit 2}) the sum of residual background and $BW$ signal distribution (Eq.~(\ref{eq:total_fit_2})).  
Finally, the uncorrected number of $K^{*}(892)^0$ mesons (for each separate $y$ and $p_T$ bin) is obtained as the integral over the $BW$ signal of \textit{total fit 2} in Figs.~\ref{fig:SM_TM_comp}, \ref{fig:SM_TM_comp_25} (bottom, right). The integral is calculated in the mass window $m_0 \pm 4\Gamma_0$.

\subsection{Uncorrected numbers of $K^{*}(892)^0$}
\label{s:Uncorrected_numbers}

\begin{figure*}
  \centering
  \includegraphics[width=0.7\textwidth]{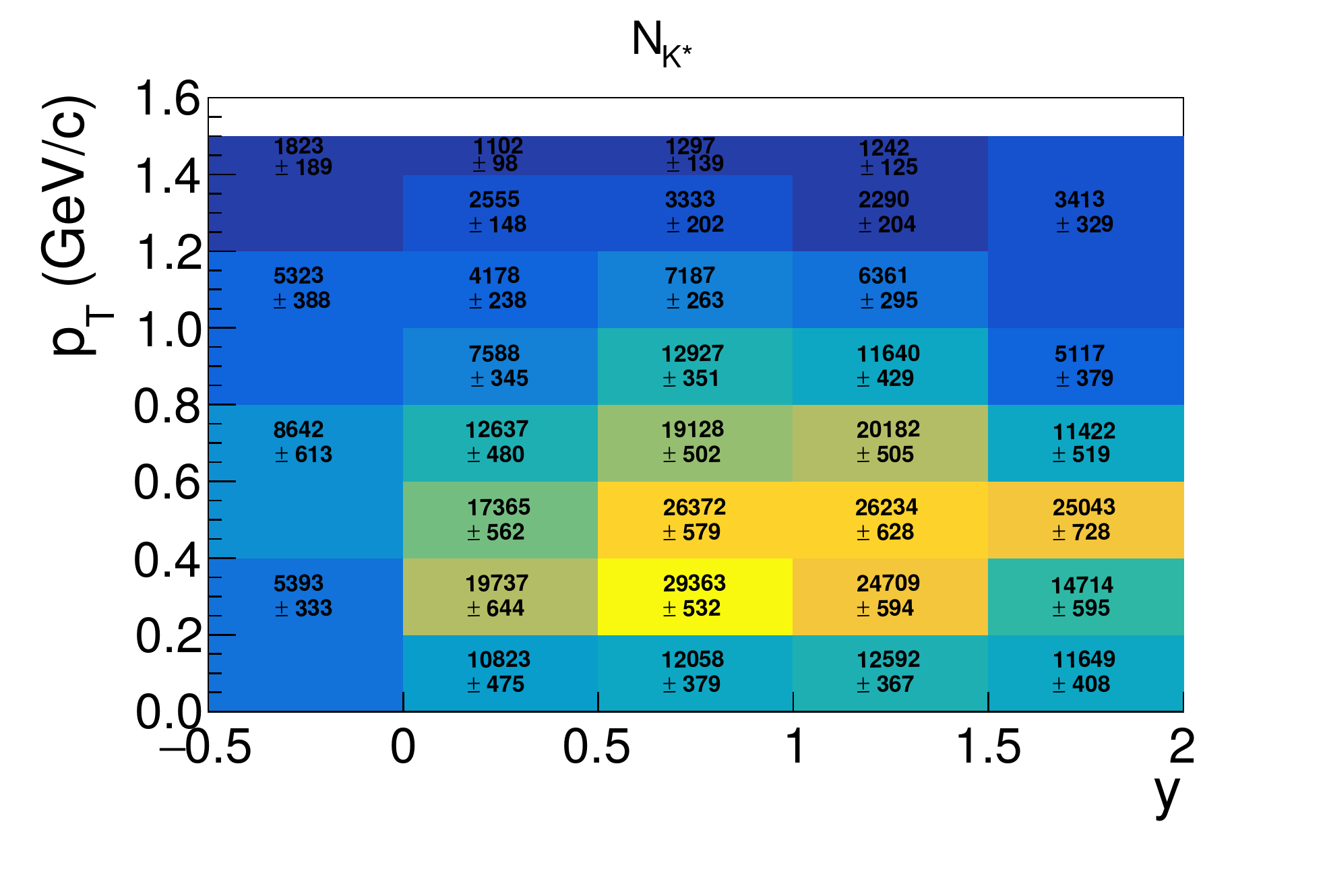}
\vspace{-0.6cm}
  \caption[]{
    (Color online) Uncorrected numbers of $K^{*}(892)^0$ obtained from the extraction procedure described in Sec.~\ref{s:signal_extraction}. The values are shown with statistical uncertainties.
}
  \label{fig:raw_yields}
\end{figure*}

Figure~\ref{fig:raw_yields} presents the uncorrected numbers of $K^{*}(892)^0$ ($N_{K^*}$) as obtained from the extraction procedure described in Sec.~\ref{s:signal_extraction}. The values are shown with statistical uncertainties.
For each $m_{K^{+}\pi^{-}}$ invariant mass bin in Fig.~\ref{fig:SM_TM_comp} or \ref{fig:SM_TM_comp_25} (bottom, right), the bin content $N_{bin}(m_{K^{+}\pi^{-}})$ was calculated as:

\begin{equation}
N_{bin}(m_{K^{+}\pi^{-}})=N_{raw}(m_{K^{+}\pi^{-}})-a \cdot T_{res}^{MC}(m_{K^{+}\pi^{-}}) - b \cdot T_{mix}^{DATA}(m_{K^{+}\pi^{-}}),
\label{eq:Nbin}
\end{equation}

where $N_{raw}(m_{K^{+}\pi^{-}})$ is the raw production in a given $m_{K^{+}\pi^{-}}$ bin, and $a$, $b$, $T_{res}^{MC}(m_{K^{+}\pi^{-}})$ and $T_{mix}^{DATA}(m_{K^{+}\pi^{-}})$ are described in Eq.~(\ref{eq:minv_function}).
The statistical uncertainty of $N_{bin}(m_{K^{+}\pi^{-}})$ can be expressed as (the notation $(m_{K^{+}\pi^{-}})$ is omitted for simplifying the formula):

\begin{equation}
\Delta N_{bin} = \sqrt {
(\Delta N_{raw})^2 + a^2 (\Delta T_{res}^{MC})^2 + b^2 (\Delta T_{mix}^{DATA})^2},
\end{equation}

where $\Delta N_{raw}$, $\Delta T_{res}^{MC}$ and $\Delta T_{mix}^{DATA}$ are the standard statistical uncertainties taken as the square root of the number of entries. For $T_{res}^{MC}$ and $T_{mix}^{DATA}$ histograms the number of entries had to be properly normalized. Due to high statistics of data, Monte Carlo, and mixed events, the uncertainties of parameters $a$ and $b$ were neglected. 
Finally, for each bin of $(y, p_T)$ in Fig.~\ref{fig:raw_yields} the uncorrected number of $K^{*}(892)^0$, $N_{K^*}(y,p_T)$, was calculated as the integral over the $BW$ signal of \textit{total fit 2} in Figs.~\ref{fig:SM_TM_comp}, \ref{fig:SM_TM_comp_25} (bottom, right). The integral was obtained within the mass window $m_0 \pm 4\Gamma_0$. The statistical uncertainty of the raw number of $K^{*}(892)^0$, $\Delta N_{K^*}(y,p_T)$, was taken as the uncertainty of the integral calculated by the ROOT~\cite{root_page} package using covariance matrix of the fitted parameters.

\subsection{Correction factors}
\label{s:Correction_factors}

In order to determine the number of $K^{*}(892)^0$ mesons produced in inelastic p+p interactions, two corrections were applied to the extracted raw number of $K^{*}(892)^0$:

\begin{itemize}
	\item [(i)] The loss of the $K^{*}(892)^0$ due to the \dedx requirement was corrected by a constant factor:
	
	\begin{equation}
		c_{dE/dx} = \frac{1}{\epsilon_{K^{+}} \cdot \epsilon_{\pi^{-}}} = 1.158,
	\end{equation}
	where $\epsilon_{K^{+}}=0.866$ and $\epsilon_{\pi^{-}}=0.997$ are the probabilities (based on the cumulative Gaussian distribution) for $K^{+}$ or $\pi^{-}$ to lie within $1.5\sigma$ or $3\sigma$ around the nominal Bethe-Bloch value.
	\item [(ii)] A detailed Monte Carlo simulation was performed to correct for geometrical acceptance, reconstruction efficiency, losses due to the trigger bias, detector acceptance as well as the quality cuts applied in the analysis. The width of the $K^{*}(892)^0$ resonance was simulated according to the known PDG value~\cite{Pierog_priv}.
The correction factors are based on $227.9 \times 10^6$ inelastic p+p events produced by the \EposLong event generator~\cite{Werner:2005jf}.
The validity of these events for calculation of the corrections was verified in Refs.~\cite{Abgrall:2013qoa, Antoni_A_PHD}.
The particles in the generated events were tracked through the \NASixtyOne apparatus using the \GeantThree package~\cite{GEANT}. The TPC response was simulated by dedicated \NASixtyOne software packages which take into account all known detector effects. The simulated events were reconstructed with the same software as used for real events and the same selection cuts were applied (except the identification cuts: \dedx and total momentum $p_{lab}$).

	For each $y$ and $p_T$ bin, the correction factor $c_{MC}(y,p_T)$ was calculated as:
	
	\begin{equation}
	c_{MC} (y,p_T) = \frac{n_{gen}(y,p_T)}{n_{sel}(y,p_T)} = \frac{N_{K^*}^{gen}(y,p_T)}{N_{events}^{gen}} /  \frac{N_{K^*}^{sel}(y,p_T)}{N_{events}^{sel}},
	\end{equation}
	where: \\
	\begin{itemize}
		\item [-] $N_{K^*}^{gen}(y,p_T)$ is the number of $K^{*}(892)^0$ generated in a given (y,\pt) bin,
		\item [-] $N_{K^*}^{sel}(y,p_T)$ is the number of $K^{*}(892)^0$ reconstructed and selected by the cuts in a given ($y, p_T$) bin. The reconstructed charged particles were matched to the simulated $K^{+}$ and $\pi^{-}$ based on cluster positions. Then the invariant mass was calculated for all $K^{+} \pi^{-}$ pairs. The reconstructed number of $K^{*}(892)^0$ was obtained by repeating the same steps (template method) as in raw data; they are described in Section~\ref{s:signal_extraction},  
		\item [-] $N_{events}^{gen}$ is the number of generated inelastic p+p interactions ($227.9 \times 10^6$),
		\item [-] $N_{events}^{sel}$ is the number of accepted p+p events ($140.1 \times 10^6$). 
	\end{itemize}
	The uncertainty of $c_{MC} (y,p_T) $ was calculated assuming that the denominator $n_{sel}(y,p_T)$ is a subset of the nominator $n_{gen}(y,p_T)$ and thus has a binomial distribution. The uncertainty of $c_{MC} (y,p_T) $ was calculated as follows:
	
	\begin{equation}
	\Delta {c_{MC} (y,p_T)} = c_{MC}(y,p_T) \sqrt{\frac{N_{K^*}^{gen}(y,p_T) - N_{K^*}^{sel}(y,p_T)}{N_{K^*}^{gen} \cdot N_{K^*}^{sel}}}
	\label{eq:dc_mc}
	\end{equation}
\end{itemize}

\begin{figure*}
  \centering
  \includegraphics[width=0.7\textwidth]{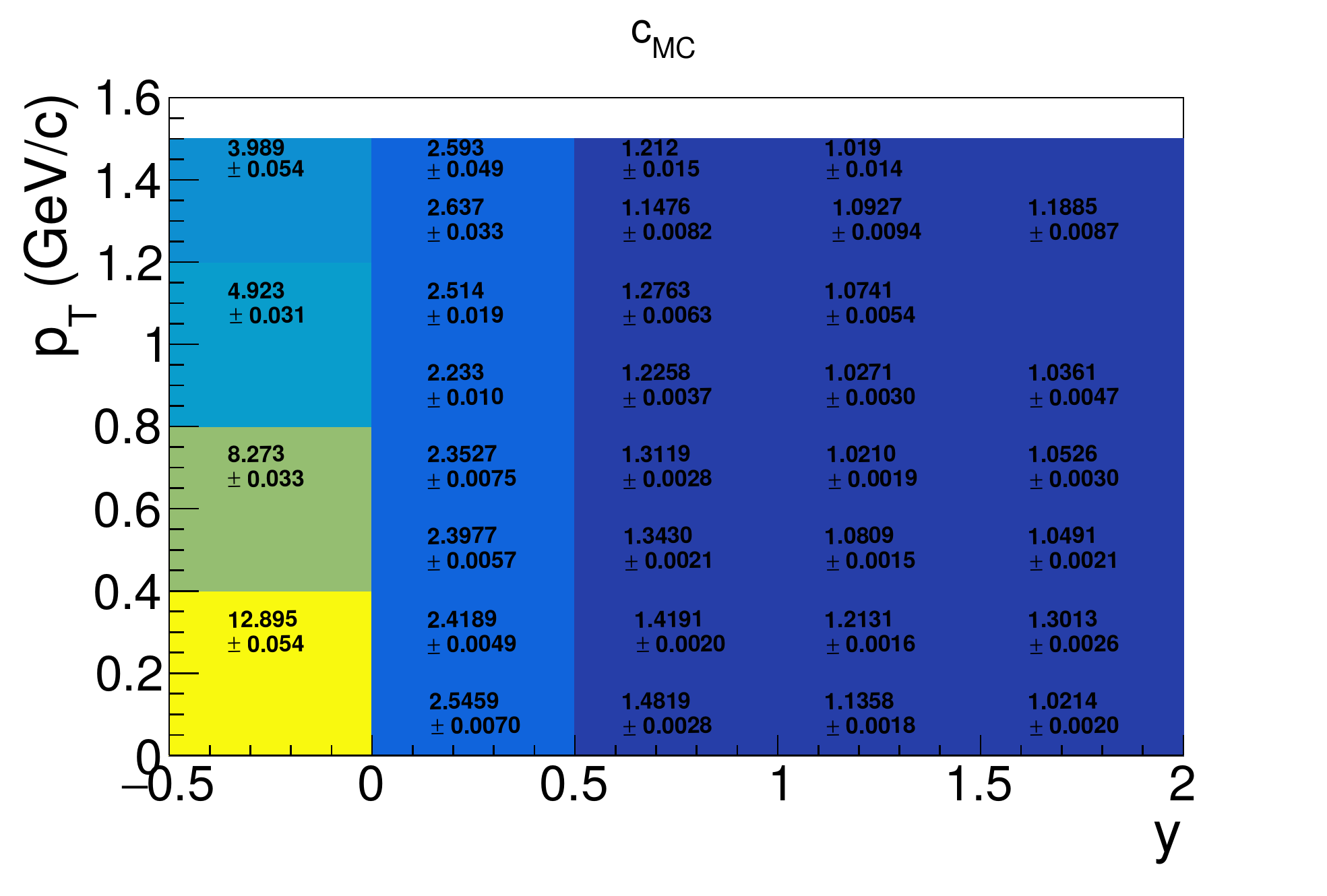}
\vspace{-0.3cm}
  \caption[]{
    (Color online) Correction factors $c_{MC}$ with statistical uncertainties. }
  \label{fig:c_MC}
\end{figure*}

The values of correction factors $c_{MC}$, together with statistical uncertainties, are presented in Fig.~\ref{fig:c_MC} for all analyzed $(y,p_T)$ bins.

\subsection{Corrected $K^{*}(892)^0$ yields}

The double-differential yield of $K^{*}(892)^0$ per inelastic event in a bin of ($y, p_T$) is calculated as follows:

\begin{equation}
\frac{d^2 n}{dy\, dp_T} (y, p_T) = \frac{1}{BR} \cdot \frac{N_{K^*}(y,p_T)}{N_{events}} \cdot \frac{c_{dE/dx} \cdot c_{MC}(y,p_T)}{\Delta y \, \Delta p_T},
\label{eq:dndydpt}
\end{equation}

where: \\
\begin{itemize}
	\item [-] $BR=2/3$ is the branching ratio of $K^{*}(892)^0$ decay into $K^{+} \pi^{-}$ pairs (obtained~\cite{Claudia_H_PHD} from the Clebsch-Gordan coefficients),
	\item [-] $N_{K^*}(y,p_T)$ is the uncorrected number of $K^{*}(892)^0$, obtained by the signal extraction procedure described in Sec.~\ref{s:signal_extraction},
	\item [-] $N_{events}$ is the number of events after cuts,
	\item [-] $c_{dE/dx}$, $ c_{MC}(y,p_T)$ are the correction factors described above,
	\item [-] $\Delta y$ and $\Delta p_T$ are the bin widths.   
\end{itemize}

The corrected double-differential yields of $K^{*}(892)^0$ together with their uncertainties are presented in Sec.~\ref{sec:results}.

\subsection{Statistical and systematic uncertainties}
\label{s:statistical_and_systematic_uncertainties}

The statistical uncertainties of the corrected double-differential yields (see Eq.~(\ref{eq:dndydpt})) take into account the statistical uncertainties of $c_{MC}(y,p_T)$ (see Eq.~(\ref{eq:dc_mc})) and the statistical uncertainties $\Delta N_{K^*} (y, p_T)$ (see Sec.~\ref{s:Uncorrected_numbers})
of the uncorrected number of $K^{*}(892)^0$. The correction $c_{dE/dx}$ has no statistical uncertainty. The final formula is expressed as follows:

\begin{equation}
	\Delta \frac{d^2 n}{dy\, dp_T} \left ( y, p_T \right) = \frac{1}{BR} \cdot \sqrt{\left ( \frac{c_{dE/dx} \cdot c_{MC}(y,p_T)}{N_{events}\, \Delta y \, \Delta p_T} \right )^2 \Delta N_{K^*}^2 (y, p_T) + \left ( \frac{N_{K^{*}}(y,p_T) \cdot c_{dE/dx}}{N_{events}\, \Delta y \, \Delta p_T} \right)^2 \Delta c_{MC}^2(y, p_T)}. 
\end{equation}

The systematic uncertainties were estimated taking into account two sources. The first group of effects is associated with the signal extraction procedure and the second with event and track quality cuts.

The considered sources of the systematic uncertainty and the corresponding modifications of the analysis method were the following: 

\begin{itemize}
	\item [(I)] The uncertainty due to the signal extraction procedure:
		\begin{itemize}
			
			\item [(i)] the lower limit of the invariant mass fitting range (see Figs.~\ref{fig:SM_TM_comp}, \ref{fig:SM_TM_comp_25} (top, right)) was changed from 0.66~\GeV to 0.69~\GeV, 
 			\item [(ii)] the initial value of the $\Gamma_{K^*}$ parameter of the signal function was changed by $\pm$8\%, 
			\item [(iii)] the initial value of the mass parameter of the Breit-Wigner distribution
			was changed by $\pm$0.3 \MeV,
			\item [(iv)] the parameters $a$, $b$ and $c$  describing the contribution of the templates in the                         fitting function (see Eq.~(\ref{eq:minv_function})) were changed by $\pm$10\%, 
			\item [(v)] the value of the $\Gamma_{K^*}$ parameter of the signal function was fixed at the PDG                         value $\Gamma_0$,
			\item [(vi)] the value of the $m_{K^*}$ parameter of the signal function was fixed at the PDG                             value $m_0$,
			\item [(vii)] in the final step of the background fit (see Figs.~\ref{fig:SM_TM_comp}, 
                        \ref{fig:SM_TM_comp_25} (bottom, right)) the standard polynomial curve of the 2nd order was 
                        changed into a polynomial curve of the 3rd order,
			\item [(viii)] the invariant mass range over which the raw number of $K^{*}(892)^0$ was integrated                        was changed from $m_0\pm 4\Gamma_0$ to $\pm 3.5\Gamma_0$ and $\pm 4.5\Gamma_0$, 
			\item [(ix)] the raw number of $K^{*}(892)^0$ was calculated as the sum of points (after 2nd order                        polynomial subtraction) instead of the $BW$ signal integral.  

		\end{itemize}

	\item [(II)] The effects of event and track quality cuts were checked by performing the analysis with the following cuts changed compared to the original values:
		\begin{itemize}
			\item [(i)] the window in which off-time beam particles are not allowed was increased from $\pm 1$ $\mu$s to $\pm1.5$  $\mu$s around the trigger particle,
	\item [(ii)] the cut on the $z$-position of the interaction vertex was changed from $[-590;-572]$ \cm to $[-591;-571]$ \cm and $[-589, -573]$ \cm, 
			\item [(iii)] the standard \dedx cuts ($\pm 3\sigma$ for $\pi^{-}$ and $\pm 1.5\sigma$ for $K^{+}$) were modified to $\pm 2.5\sigma$ for $\pi^{-}$, $\pm 1.0\sigma$ for $K^{+}$ (narrower cut) and $\pm 3.5\sigma$ for $\pi^{-}$, $\pm 2.0\sigma$ for $K^{+}$ (wider cut),
			\item [(iv)] the minimum required total number of points in all TPCs for $K^{*}(892)^0$ decay products was changed from 30 to 25 and 35,
			\item [(v)] the minimum required number of clusters in both VTPCs for $K^{*}(892)^0$ decay products was changed from 15 to 12 and 18,
			\item [(vi)] the impact parameter cuts for the tracks were turned off. 
		\end{itemize} 
\end{itemize}

For each of the possible sources described above the partial systematic uncertainty $\sigma_i$ was calculated as half of the difference between the lowest and the highest value obtained by varying the given parameter. Then, the final systematic uncertainty was taken as: 
\begin{math}
\sigma_{sys} = \sqrt{\sum \sigma_i^2}
\end{math}.   
The contributions of uncertainties $\sigma_i$ to the total uncertainty are negligible for I (ii), I (iii), and I (iv).  
The final systematic uncertainties are shown in the figures as light red shaded bands.

%%% Local Variables: 
%%% mode: latex
%%% TeX-master: "main"
%%% End: 

\section{Results}\label{sec:results}

%%%%%%%%%%%%%%%%%%%%%%%%%%%%%%
\subsection{Mass and width of $K^{*}(892)^0$}

\begin{figure*}
	\centering
	\includegraphics[width=0.49\textwidth]{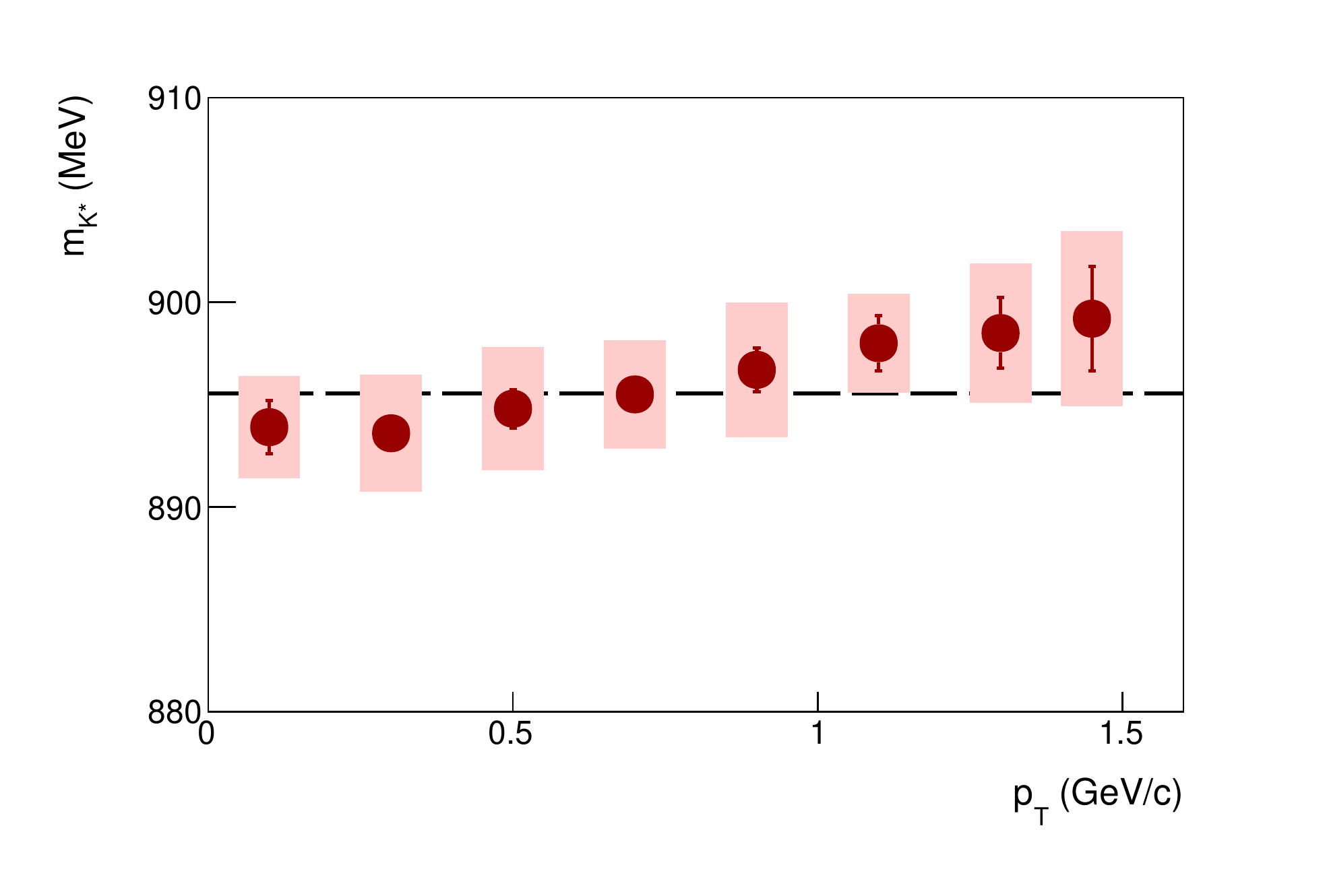}
	\includegraphics[width=0.49\textwidth]{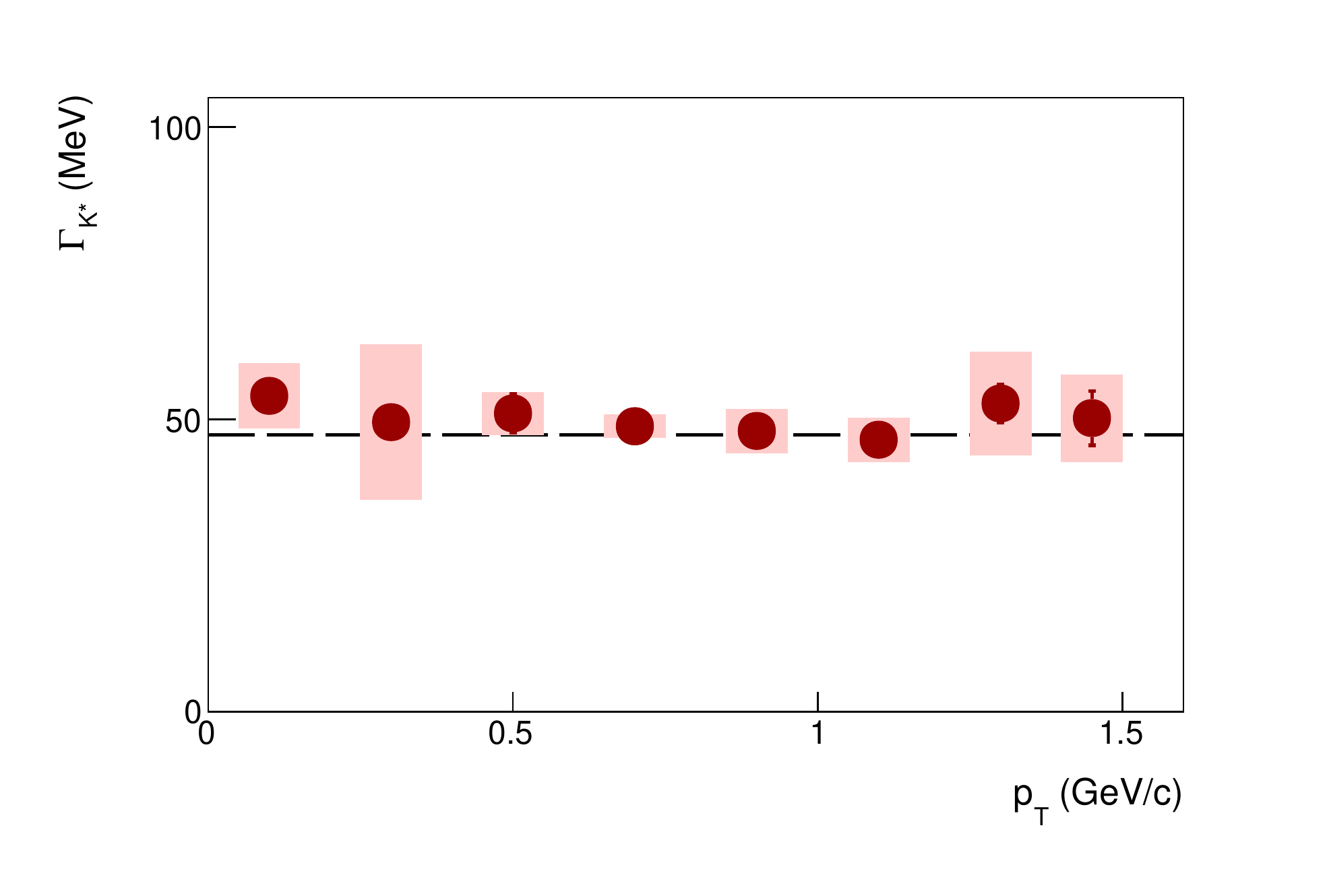}
\vspace{-0.3cm}
	\caption{(Color online) The transverse momentum dependence of mass and width of $K^{*}(892)^0$ mesons fitted for $0 < y < 0.5$. The numerical data are listed in Table~\ref{tab:mass_width}. The horizontal lines represent PDG values $m_0=895.55$~\MeV and $\Gamma_0=47.3$~\MeV~\cite{PDG}.}
\label{fig:mass_gamma_NA61only}
\end{figure*}

The values of mass and width of $K^{*}(892)^0$ mesons were extracted from the fits to background subtracted invariant mass spectra (see Sec.~\ref{s:signal_extraction}). They are presented in Fig.~\ref{fig:mass_gamma_NA61only} in different transverse momentum bins (numerical data are listed in Table~\ref{tab:mass_width}). The results are shown for the rapidity range $0 < y < 0.5$. 
Within uncertainties, the $\Gamma_{K^*}$ values are consistent with information provided by the PDG. However, one observes a slight increase of the $m_{K^*}$ parameter with \pt with an average close to the PDG value. The corresponding slope is significant since a large part of the shown systematic uncertainty is due to the magnetic field uncertainty (see below). The points (with their statistical uncertainties), presented in the left panel of Fig.~\ref{fig:mass_gamma_NA61only}, were fitted with a linear function resulting in the slope parameter value equal to 4.5 $\pm$ 1.2.  
The change of the $m_{K^*}$ parameter with transverse momentum does not introduce a systematic variation of the $K^{*}(892)^0$ yield since the parameter is fitted in each ($y$, \pt) bin, and the signal integration range ($\approx$ 380~\MeV) is much larger than the $m_{K^*}$ change ($\approx$ 6~\MeV).

The magnetic field strength was verified with a precision of better than 1\% by studying the $K^0_S$ and $\Lambda$ invariant mass distributions~\cite{SR_2008}.
In order to check how the magnetic field calibration influences the results, the momentum components of $K^{*}(892)^0$ decay products (kaons and pions) were varied by $\pm 1\%$. 
   Such a change did not affect $K^{*}(892)^0$ width and yield significantly. However, the resulting changes of the mass parameter are equal or larger than uncertainties described in Sec.~\ref{s:statistical_and_systematic_uncertainties}, and they were taken into account in the calculation of the final uncertainty of the $K^{*}(892)^0$ mass parameter shown in Fig.~\ref{fig:mass_gamma_NA61only} (left) and Table~\ref{tab:mass_width}. 

The comparison of mass and width of $K^{*}(892)^0$ mesons with other experiments is shown in Sec.~\ref{sec:comparison}.

\begin{table}
\centering
\begin{tabular}{|c|c|c|}
\hline
$p_T$ (\GeVc) & $m_{K^*}$ (MeV) & $\Gamma_{K^*}$ (MeV) \\
\hline
(0.0;0.2) & $893.9 \pm 1.3 \pm 2.5$ & $54.00 \pm 0.64 \pm 5.5$  \\
\hline
(0.2;0.4) & $893.63 \pm 0.90 \pm 2.8$ & $49.50 \pm 0.93 \pm 13$  \\
\hline
(0.4;0.6) & $894.81 \pm 0.93 \pm 3.0$ & $50.9 \pm 3.3 \pm 3.6$  \\
\hline
(0.6;0.8) & $895.50 \pm 0.92 \pm 2.6$ & $48.8 \pm 1.8 \pm 2.0$  \\
\hline
(0.8;1.0) & $896.7 \pm 1.1 \pm 3.3$ & $48.0 \pm 2.1 \pm 3.7$  \\
\hline
(1.0;1.2) & $898.0 \pm 1.4 \pm 2.4$ & $46.5 \pm 2.5 \pm 3.7$  \\
\hline
(1.2;1.4) & $898.5 \pm 1.7 \pm 3.4$ & $52.7 \pm 3.3 \pm 8.8$  \\
\hline
(1.4;1.5) & $899.2 \pm 2.6 \pm 4.3$ & $50.2 \pm 4.6 \pm 7.4$  \\
\hline
\end{tabular}
\caption{Numerical values of mass and width of $K^{*}(892)^0$ mesons fitted in $0 < y < 0.5$ and presented in Fig.~\ref{fig:mass_gamma_NA61only}. The first uncertainty is statistical, while the second one is systematic.}
\label{tab:mass_width}
\end{table}

%%%%%%%%%%%%%%%%%%%%%%%%%%%%%%

\subsection{Double-differential $K^{*}(892)^0$ spectra}

\begin{figure*}
	\centering
	\includegraphics[width=0.7\textwidth]{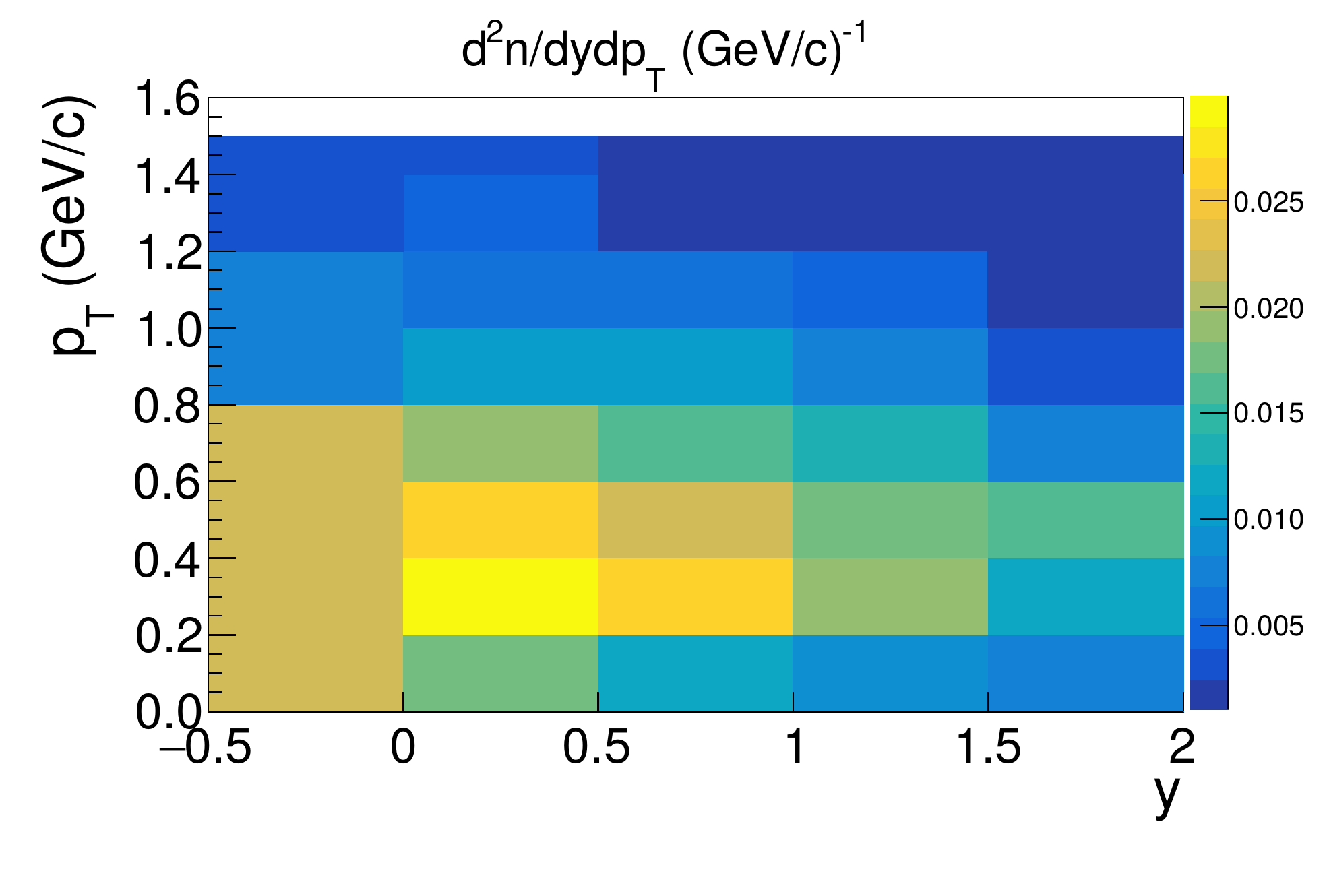}
\vspace{-0.5cm}
	\caption[]{(Color online) Double-differential $K^{*}(892)^0$ spectra in inelastic p+p interaction at 158~\GeVc in bins of ($y, p_T$) as obtained from Eq.~(\ref{eq:dndydpt}). 
The numerical values are given in Table~\ref{tab:dndydpt}.
}
	\label{fig:dndydpt_2D}
\end{figure*}

The double-differential yields of $K^{*}(892)^0$ mesons
in inelastic p+p interaction at 158~\GeVc in bins of ($y, p_T$) are presented in Fig.~\ref{fig:dndydpt_2D}. 
The numerical values with statistical and systematic uncertainties are presented in Table~\ref{tab:dndydpt}.

%%%%%%%%%%%%%%%%%%%%%%%%%%%%%%

\subsection{Transverse momentum and transverse mass spectra}

\begin{figure*}
	\centering
	\includegraphics[width=0.7\textwidth]{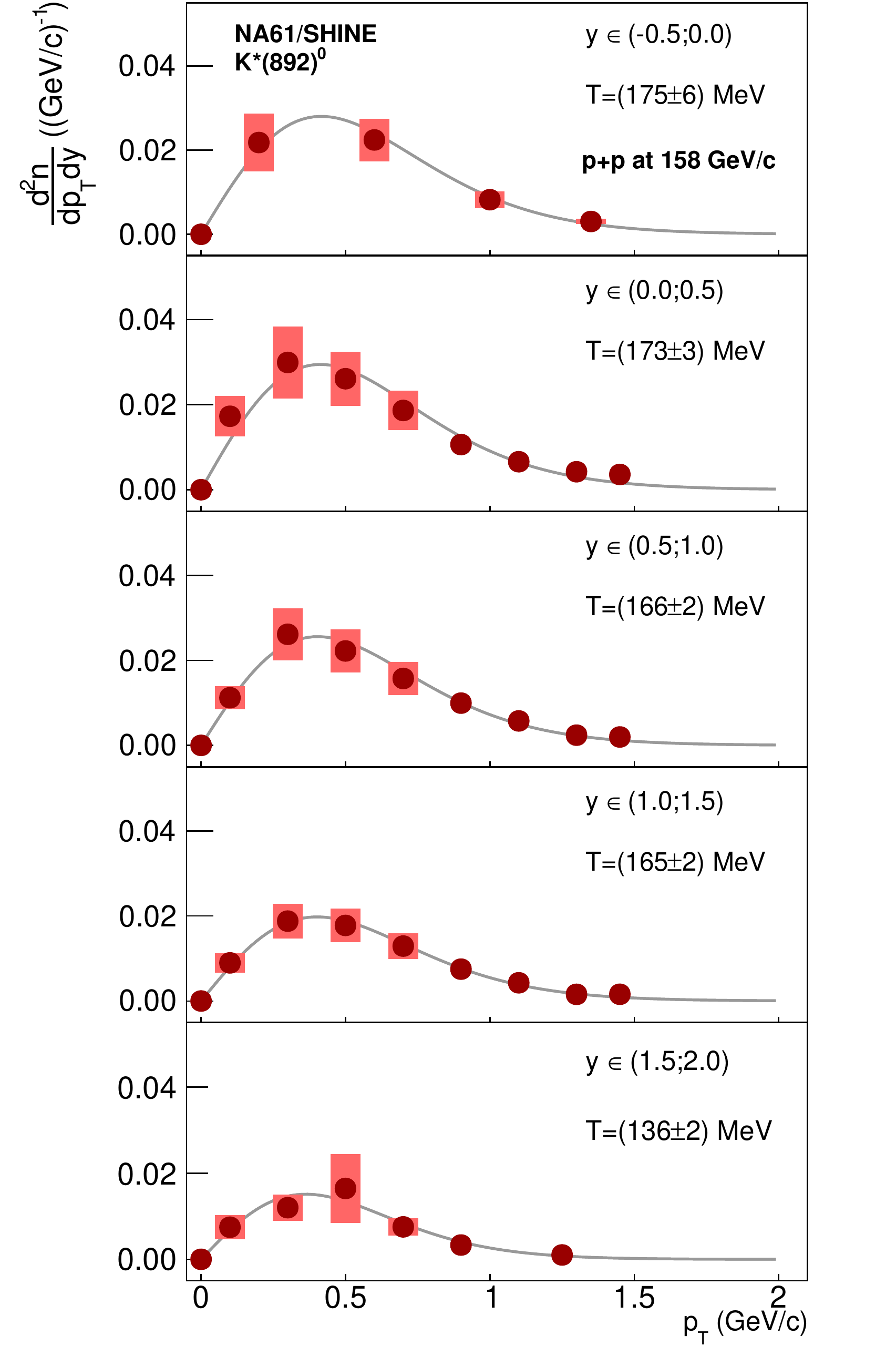}
	\caption[]{(Color online) Transverse momentum spectra $\frac{d^2 n}{dy\, dp_T}$ for five bins of rapidity. The fitted function (solid line) is given by Eq.~(\ref{eq:fit_to_dndydpt}). The numerical values are listed in Table~\ref{tab:dndydpt} and the fitted inverse slope parameters $T$ for each bin are given in the legends.  }
	\label{fig:dndydpt}
\end{figure*}

Figure~\ref{fig:dndydpt} shows the double-differential yields of $K^{*}(892)^0$ mesons as function of \pt presented for separate rapidity bins. The corresponding numerical values are listed in Table~\ref{tab:dndydpt}.  

In order to measure the inverse slope parameter $T$ of transverse momentum spectra and to estimate the yield of $K^{*}(892)^0$ mesons in the unmeasured high \pt region, the function:
\begin{equation}
f(p_T) = A \cdot p_T \, \exp \left ( - \frac{\sqrt{p_T^2 + m_0^2}}{T} \right)
\label{eq:fit_to_dndydpt}
\end{equation}
was fitted to the measurements shown in Fig.~\ref{fig:dndydpt}. 
The inverse slope parameters obtained from the fits are cited in the figure legends.

	\begin{table}
\centering
%		\small
		\begin{tabular}{|c|c|c|c|}
			\hline
			& \multicolumn{3}{c|}{$y$} \\
			\cline{2-4}
			$p_T$ (\GeVc) &  (-0.5;0.0) &  (0.0;0.5) &  (0.5;1.0) \\
			\hline
			(0.0;0.2) & \multirow{2}{*}{ 21.8 $\pm$ 1.3 $\pm$ 6.7} &  17.28 $\pm$ 0.76 $\pm$ 4.7 &  11.21 $\pm$ 0.35 $\pm$ 2.6 \\
			\cline{1-1}
			\cline{3-4}
			(0.2;0.4) & &  29.94 $\pm$ 0.98 $\pm$ 8.4 &  26.13 $\pm$ 0.47 $\pm$ 6.0 \\
			\hline
			(0.4;0.6) & \multirow{2}{*}{ 22.4 $\pm$ 1.6 $\pm$ 5.0} &  26.11 $\pm$ 0.85 $\pm$ 6.3 &  22.21 $\pm$ 0.49 $\pm$ 4.9 \\
			\cline{1-1}
			\cline{3-4}
			(0.6;0.8) & &  18.65 $\pm$ 0.71 $\pm$ 4.6 &  15.74 $\pm$ 0.41 $\pm$ 3.8 \\
			\hline
			(0.8;1.0) & \multirow{2}{*}{ 8.22 $\pm$ 0.60 $\pm$ 1.9} &  10.62 $\pm$ 0.48 $\pm$ 2.6 &  9.94 $\pm$ 0.27 $\pm$ 2.3 \\
			\cline{1-1}
			\cline{3-4}
			(1.0;1.2) & &  6.59 $\pm$ 0.38 $\pm$ 1.8 &  5.75 $\pm$ 0.21 $\pm$ 1.4  \\
			\hline
			(1.2;1.4) & \multirow{2}{*}{ 3.04 $\pm$ 0.32 $\pm$ 0.51 } &  4.22 $\pm$ 0.25 $\pm$ 1.2 &  2.39 $\pm$ 0.15 $\pm$ 0.56 \\
			\cline{1-1}
			\cline{3-4}
			(1.4;1.5) & &  3.58 $\pm$ 0.33 $\pm$ 1.2 &  1.97 $\pm$ 0.21 $\pm$ 0.55\\
			\hline
			& \multicolumn{2}{c|}{$y$} & \\
			\cline{2-3}
			$p_T$ (\GeVc) &   (1.0;1.5) &  (1.5;2.0) &\\
			\cline{1-3}
			(0.0;0.2) &  8.97 $\pm$ 0.26 $\pm$ 2.2 &  7.46 $\pm$ 0.26 $\pm$ 2.7 &\\
			\cline{1-3}
			(0.2;0.4) &  18.79 $\pm$ 0.45 $\pm$ 4.0 &  12.00 $\pm$ 0.49 $\pm$ 3.0 &\\
			\cline{1-3}
			(0.4;0.6) &  17.78 $\pm$ 0.43 $\pm$ 3.9 &  16.47 $\pm$ 0.48 $\pm$ 7.9 &\\
			\cline{1-3}
			(0.6;0.8) &  12.92 $\pm$ 0.32 $\pm$ 3.0 &  7.54 $\pm$ 0.34 $\pm$ 1.9 &\\
			\cline{1-3}
			(0.8;1.0) &  7.49 $\pm$ 0.28 $\pm$ 1.7 &  3.32 $\pm$ 0.25 $\pm$ 0.47 &\\
			\cline{1-3}
			(1.0;1.2) &  4.28 $\pm$ 0.20 $\pm$ 1.1 &  \multirow{3}{*}{1.017 $\pm$ 0.082 $\pm$ 0.38} &\\
			\cline{1-2}
			(1.2;1.4) &  1.57 $\pm$ 0.14 $\pm$ 0.44 & &\\
			\cline{1-2}
			(1.4;1.5) &  1.59 $\pm$ 0.16 $\pm$ 0.68 & &\\
			\hline
		\end{tabular}
	\caption{Numerical values of double-differential yields $\frac{d^2 n}{dy\, dp_T}$ presented in Fig.~\ref{fig:dndydpt}, given in units of $10^{-3}$ (\GeVc)$^{-1}$. The first uncertainty is statistical, while the second one is systematic.}
	\label{tab:dndydpt}
	\end{table}

%%%%%%%%%%%%%%%%%%%%%%%%%%%%%%

\begin{figure*}
	\centering
	\includegraphics[width=0.7\textwidth]{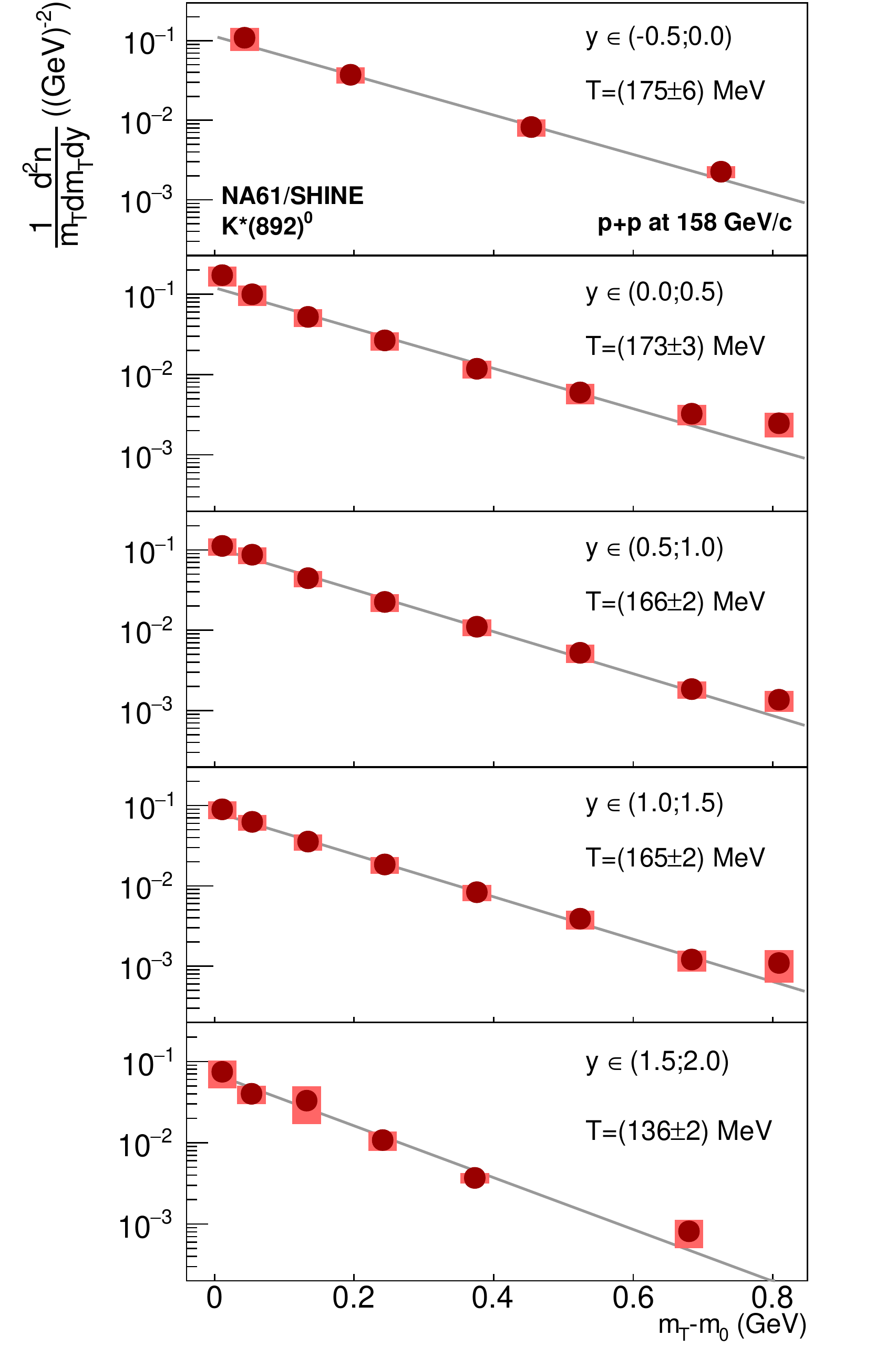}
	\caption[]{(Color online) Transverse mass spectra $\frac{1}{m_T}\frac{d^2 n}{dm_T\, dy}$ for five bins of rapidity. The numerical values are listed in Table~\ref{tab:dndydmt}. The solid lines represent function given by Eqs.~(\ref{eq:fit_to_dndydpt}) and (\ref{eq:dndydmt}) with $A$ and $T$ parameters taken from Fig.~\ref{fig:dndydpt}. } 
	\label{fig:dndydmt}
\end{figure*}

The transverse mass ($m_T \equiv \sqrt{p_T^2 + m_0^2}$) spectra $\frac{1}{m_T}\frac{d^2 n}{dm_T dy}$ were calculated based on $\frac{d^2 n}{dy dp_T}$ spectra according to:
\begin{equation}
\frac{1}{m_T}\frac{d^2 n}{dm_T\, dy} = \frac{1}{p_T}\frac{d^2 n}{dy\, dp_T}.
\label{eq:dndydmt}
\end{equation} 
The results are shown in Fig.~\ref{fig:dndydmt} and the numerical values are presented in Table~\ref{tab:dndydmt}.

	\begin{table}
\centering
%		\small
		\begin{tabular}{|c|c|c|c|c|}
			\hline
			& & \multicolumn{3}{c|}{$y$} \\
			\cline{3-5}
			\makecell{$m_T - m_0$ \\ (GeV)} & $p_T$ (\GeVc) & (-0.5;0.0) & (0.0;0.5) & (0.5;1.0)  \\
			\hline
			0.011 & (0.0;0.2) & \multirow{2}{*}{ 109.0 $\pm$ 6.7 $\pm$ 34} &  172.8 $\pm$ 7.6 $\pm$  47 &  112.1 $\pm$ 3.5 $\pm$  26 \\
			\cline{1-2} 
			\cline{4-5}
			0.054 & (0.2;0.4) & &  99.8 $\pm$ 3.3 $\pm$ 28 &  87.1 $\pm$ 1.6  $\pm$  20\\
			\hline
			0.134 & (0.4;0.6) & \multirow{2}{*}{ 37.4 $\pm$ 2.6 $\pm$ 8.3} &  52.2 $\pm$ 1.7 $\pm$ 13 &  44.42 $\pm$ 0.97 $\pm$ 9.9 \\
			\cline{1-2}
			\cline{4-5}
			0.244 & (0.6;0.8) & &  26.6 $\pm$ 1.0 $\pm$  6.5 & 22.48 $\pm$ 0.59 $\pm$  5.5\\
			\hline
			0.376 & (0.8;1.0) & \multirow{2}{*}{  8.21 $\pm$ 0.60 $\pm$ 1.9} & 11.81 $\pm$ 0.54  $\pm$ 2.9 &  11.04 $\pm$ 0.30 $\pm$  2.5 \\
			\cline{1-2}
			\cline{4-5}
			0.524 & (1.0;1.2) & &  5.99 $\pm$ 0.34 $\pm$ 1.6 &  5.23 $\pm$ 0.19  $\pm$ 1.3\\
			\hline
			0.684 & (1.2;1.4) & \multirow{2}{*}{ 2.25 $\pm$ 0.23 $\pm$ 0.38} &  3.25 $\pm$ 0.19 $\pm$ 0.91 &  1.85 $\pm$ 0.11 $\pm$ 0.43 \\
			\cline{1-2}
			\cline{4-5}
			0.809 & (1.4;1.5) & &  2.47 $\pm$  0.22 $\pm$ 0.81 & 1.36 $\pm$ 0.15 $\pm$ 0.38 \\
			\hline
			& & \multicolumn{2}{c|}{$y$} & \\
			\cline{3-4}
			\makecell{$m_T - m_0$ \\ (GeV)} & $p_T$ (\GeVc) &   (1.0;1.5) &  (1.5;2.0) &\\
			\cline{1-4}
			0.011 & (0.0;0.2) &  89.7 $\pm$ 2.6 $\pm$ 22 &  74.6 $\pm$ 2.6 $\pm$ 27 & \\
			\cline{1-4}
			0.054 & (0.2;0.4) &   62.6 $\pm$ 1.5 $\pm$ 13 &  40.0 $\pm$ 1.6  $\pm$ 9.9 & \\
			\cline{1-4}
			0.134 & (0.4;0.6) &  35.57 $\pm$ 0.85 $\pm$ 7.8 &  32.95 $\pm$ 0.96 $\pm$  16 &\\
			\cline{1-4}
			0.244 & (0.6;0.8) &  18.46  $\pm$ 0.46 $\pm$ 4.3 & 10.77 $\pm$ 0.49 $\pm$ 2.8 & \\
			\cline{1-4}
			0.376 & (0.8;1.0) &   8.33 $\pm$ 0.31 $\pm$ 1.8 & 3.69 $\pm$ 0.27 $\pm$ 0.52 & \\
			\cline{1-4}
			0.524 & (1.0;1.2) &  3.89 $\pm$ 0.18 $\pm$ 0.99 &  \multirow{3}{*}{0.814 $\pm$ 0.065 $\pm$ 0.31} & \\
			\cline{1-3}
			0.684 & (1.2;1.4) &  1.21 $\pm$ 0.11  $\pm$ 0.34 &  & \\
			\cline{1-3}
			0.809 & (1.4;1.5) &  1.09 $\pm$ 0.11 $\pm$ 0.47 & & \\
			\hline
		\end{tabular}
	\caption{Numerical values of double-differential yields $\frac{1}{m_T}\frac{d^2 n}{dm_T\, dy}$ given in units of $10^{-3}$ (\GeV)$^{-2}$ and presented in Fig.~\ref{fig:dndydmt}; the values of $m_T - m_0$ specify the bin centers. The first uncertainty is statistical, while the second one is systematic. 
}
	\label{tab:dndydmt}
	\end{table}

For the mid-rapidity region ($0 < y < 0.5$) the inverse slope parameter of the transverse momentum spectrum was found to be equal to $T=(173 \pm 3 \pm 9)$~\MeV, where statistical uncertainty (the first one) is equal to the uncertainty of the fit parameter, and the systematic uncertainty was estimated in the way described in Sec.~\ref{s:statistical_and_systematic_uncertainties}. The NA49 experiment measured the $T$ parameter of the \pt spectrum in the rapidity range $0.2 < y < 0.7$ and reported a value $T=(166 \pm 11 \pm 10)$~\MeV~\cite{Anticic:2011zr}.

%%%%%%%%%%%%%%%%%%%%%%%%%%%%%%

\subsection{$p_T$-integrated and extrapolated rapidity distribution}

The rapidity distribution $\frac{dn}{dy}$ was calculated by integrating and extrapolating (for the non-measured high-\pt region) the $\frac{d^2 n}{dy \, dp_T}$ spectrum:

\begin{equation}
\frac{dn}{dy} = \sum_i \frac{d^2 n}{dy \, dp_T} \cdot dp_T + \frac{A_{p_T}}{I_{p_T}} \sum_i  \frac{d^2 n}{dy \, dp_T} \cdot dp_T,
\end{equation} 

where: \\
\begin{equation}
A_{p_T} = \int_{1.5}^{+ \infty} A \cdot p_T \, \exp \left ( - \frac{\sqrt{p_T^2 + m_0^2}}{T} \right) dp_T, \hspace{0.7cm} I_{p_T} = \int_{0}^{1.5} A \cdot p_T \, \exp \left ( - \frac{\sqrt{p_T^2 + m_0^2}}{T} \right) dp_T. 
\end{equation}

The parameters $T$ were taken from the corresponding plots in Fig.~\ref{fig:dndydpt}.
The statistical uncertainties of \pt-integrated and extrapolated $\frac{dn}{dy}$ values were calculated as follows:

\begin{equation}
\Delta \frac{dn}{dy} = \sqrt{\left ( 1 + \frac{A_{p_T}}{I_{p_T}} \right )^2 \cdot \sum_i dp_T^2 \cdot \left ( \Delta \frac{d^2 n}{dy \, dp_T} \right )^2}. 
\end{equation}

\begin{figure*}
	\centering
	\includegraphics[width=0.9\textwidth]{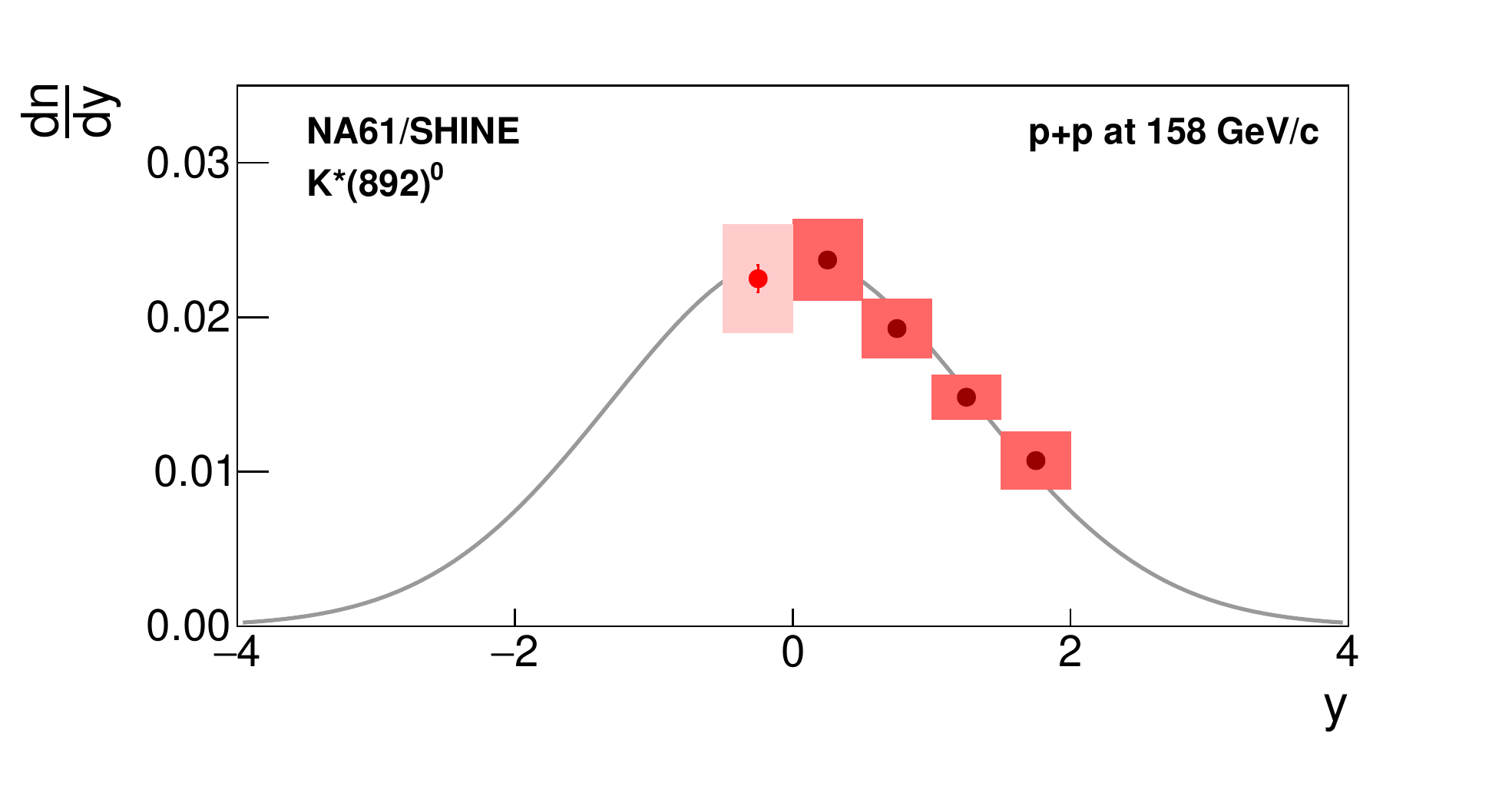}
\vspace{-0.5cm}
	\caption[]{(Color online) The \pt-integrated and extrapolated rapidity distribution. The fitted Gaussian function (solid line) is given by Eq.~(\ref{eq:fit_to_dndy}); the first point (with $y < 0$) was not included in the fit (see the text for details). The numerical data are listed in Table~\ref{tab:dndy}.   }
	\label{fig:dndy}
\end{figure*}

The \pt-integrated and extrapolated $\frac{dn}{dy}$ spectrum of $K^{*}(892)^0$ mesons is plotted in Fig.~\ref{fig:dndy} and the numerical values are listed in Table~\ref{tab:dndy}.

A Gaussian function:
\begin{equation}
f(y) = A \cdot \exp \left ( - \frac{y^2}{2\, \sigma_y^2} \right)
\label{eq:fit_to_dndy}
\end{equation}
was fitted to the data points to measure the width $\sigma_y$ of the $K^{*}(892)^0$ rapidity distribution. The first point with $y < 0$ is plotted only to check the symmetry of the distribution and was not included in the fit.
The fit was also used to determine the mean multiplicity $\langle K^{*}(892)^0 \rangle$ (see Sec.~\ref{sec:mean_multip} for details of the procedure). The statistical uncertainty of $\sigma_y$ was taken from the fit and the systematic uncertainty was estimated in the way described in Sec.~\ref{s:statistical_and_systematic_uncertainties}. The numerical values of $\sigma_y$ and $\langle K^{*}(892)^0 \rangle$ are listed in Table~\ref{tab:dndy}.

	\begin{table}
		\centering
		\begin{tabular}{|c|c|}
			\hline
			$y$ & $\frac{dn}{dy}$ \\
			\hline
			(-0.5;0.0) & (22.50 $\pm$ 0.89 $\pm$ 3.5) $\cdot 10^{-3}$ \\
			\hline
			(0.0;0.5) & (23.71 $\pm$ 0.37 $\pm$ 2.6) $\cdot 10^{-3}$ \\
			\hline
			(0.5;1.0) & (19.27 $\pm$ 0.20 $\pm$ 1.9) $\cdot 10^{-3}$ \\
			\hline
			(1.0;1.5) & (14.83 $\pm$ 0.17 $\pm$ 1.4) $\cdot 10^{-3}$ \\
			\hline
			(1.5;2.0) & (10.73 $\pm$ 0.22 $\pm$ 1.8) $\cdot 10^{-3}$ \\
			\hline
			\hline
			$\sigma_y$ & 1.31 $\pm$ 0.15 $\pm$ 0.09 \\
			\hline
			$\langle K^{*} (892)^0 \rangle$ & (78.44 $\pm$ 0.38 $\pm$ 6.0) $\cdot 10^{-3}$\\  
			\hline
		\end{tabular}
	\caption{Numerical values of the \pt-integrated and extrapolated $\frac{dn}{dy}$ distribution presented in Fig.~\ref{fig:dndy}. The first uncertainty is statistical, while the second one is systematic. Additionally, the width of the Gaussian fit to the $\frac{dn}{dy}$ distribution, as well as the mean multiplicity of $K^{*}(892)^0$ mesons are shown (see the text for details). }
	\label{tab:dndy}
\end{table}

%%%%%%%%%%%%%%%%%%%%%%%%%%%%%%

\subsection{Mean multiplicity of $K^{*}(892)^0$}
\label{sec:mean_multip}

The mean multiplicity of $K^*(892)^0$ mesons was calculated as the sum of measured points in Fig.~\ref{fig:dndy} (the first point, with $y < 0$, was not included in the sum) and the integral of the fitted Gaussian function Eq.~(\ref{eq:fit_to_dndy}) in the unmeasured region assuming symmetry around $y = 0$:
\begin{equation}
\langle K^*(892)^0 \rangle = \sum_{i}^{} \frac{dn}{dy} \cdot dy + \left ( \frac{A_{y-}+A_{y+}}{I_y} \right ) \sum_{i}^{} \frac{dn}{dy} \cdot dy,
\end{equation}
where: \\
\begin{equation}
A_{y-} = \int_{-\infty}^{0} A \cdot e^{- \frac{y^2}{2 \sigma_y^2}}dy, \hspace{0.7cm}
A_{y+} = \int_{2.0}^{+\infty} A \cdot e^{- \frac{y^2}{2 \sigma_y^2}}dy, \hspace{0.7cm}
I_{y} = \int_{0}^{2.0} A \cdot e^{- \frac{y^2}{2 \sigma_y^2}}dy.
\end{equation}

The statistical uncertainty of $\langle K^{*}(892)^0 \rangle$ was obtained from the formula:
\begin{equation}
\Delta \langle K^{*}(892)^0 \rangle = \sqrt{\left ( 1 + \frac{A_{y-}+A_{y+}}{I_{y}} \right )^2 \cdot \sum_{i} dy^2 \cdot \left ( \Delta \frac{dn}{dy} \right )^2},
\end{equation}
and the systematic uncertainty was estimated in the way described in Sec.~\ref{s:statistical_and_systematic_uncertainties}.

The mean multiplicity of $K^*(892)^0$ mesons, produced in inelastic p+p collisions at 158~\GeVc, is equal to (78.44 $\pm$ 0.38 $\pm$ 6.0) $\cdot 10^{-3}$, where the first uncertainty is statistical and the second one is systematic.

%%% Local Variables: 
%%% mode: latex
%%% TeX-master: "main"
%%% End: 

\section{Comparison with world data and model predictions}\label{sec:comparison}

%%%%%%%%%%%%%%%%%%%%%%%%%

This section compares the \NASixtyOne measurements in inelastic p+p interactions at 158~\GeVc with publicly available world data as well as with predictions from microscopic and statistical models.

%%%%%%%%%%%%%%%%%%%%%%%%%%%%%%%%%%%%%%%%%%%   

\subsection{Mass and width of $K^{*}(892)^0$}

\begin{figure*}
	\centering
	\includegraphics[width=1.0\textwidth]{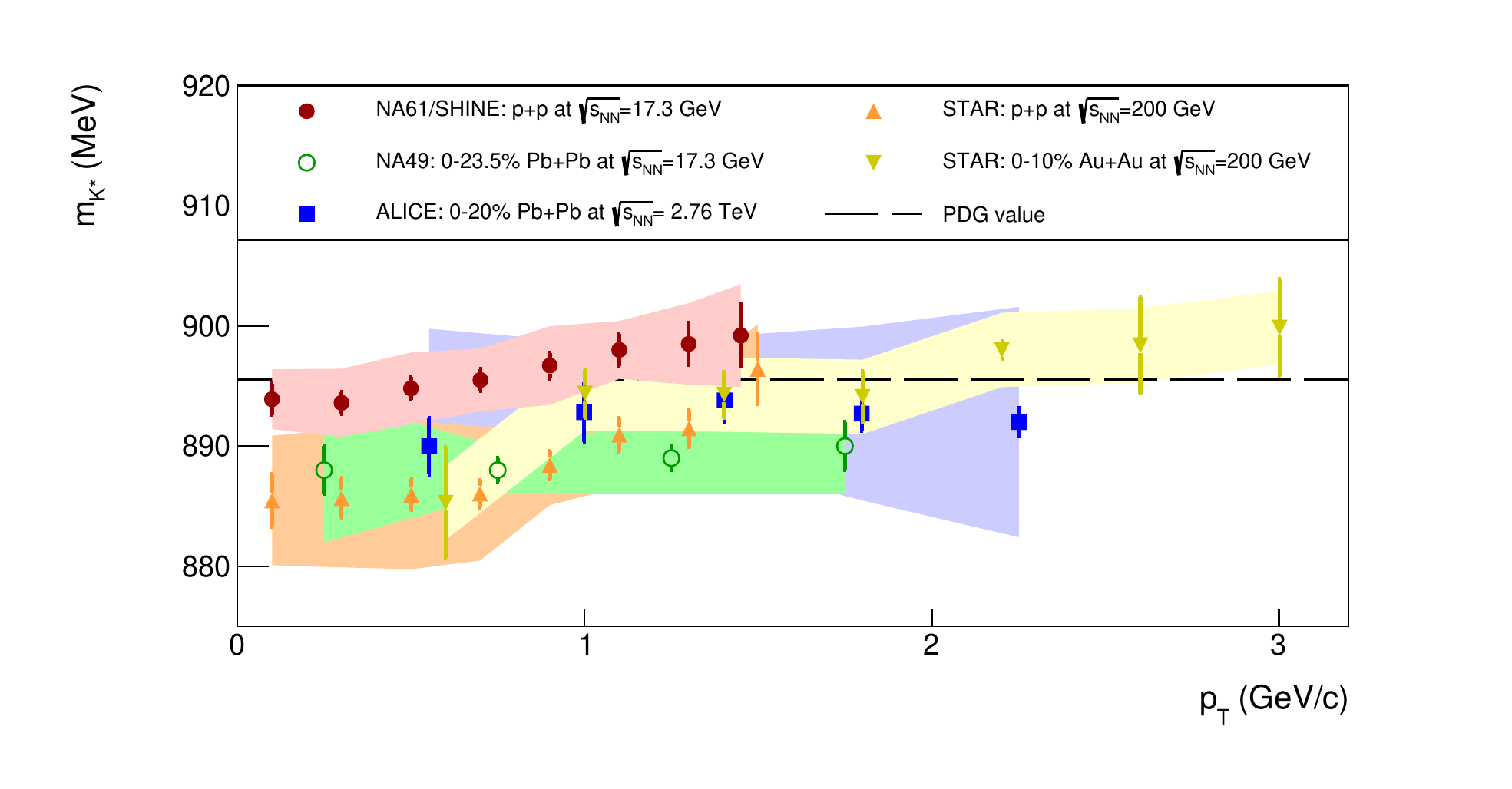}
	\includegraphics[width=1.0\textwidth]{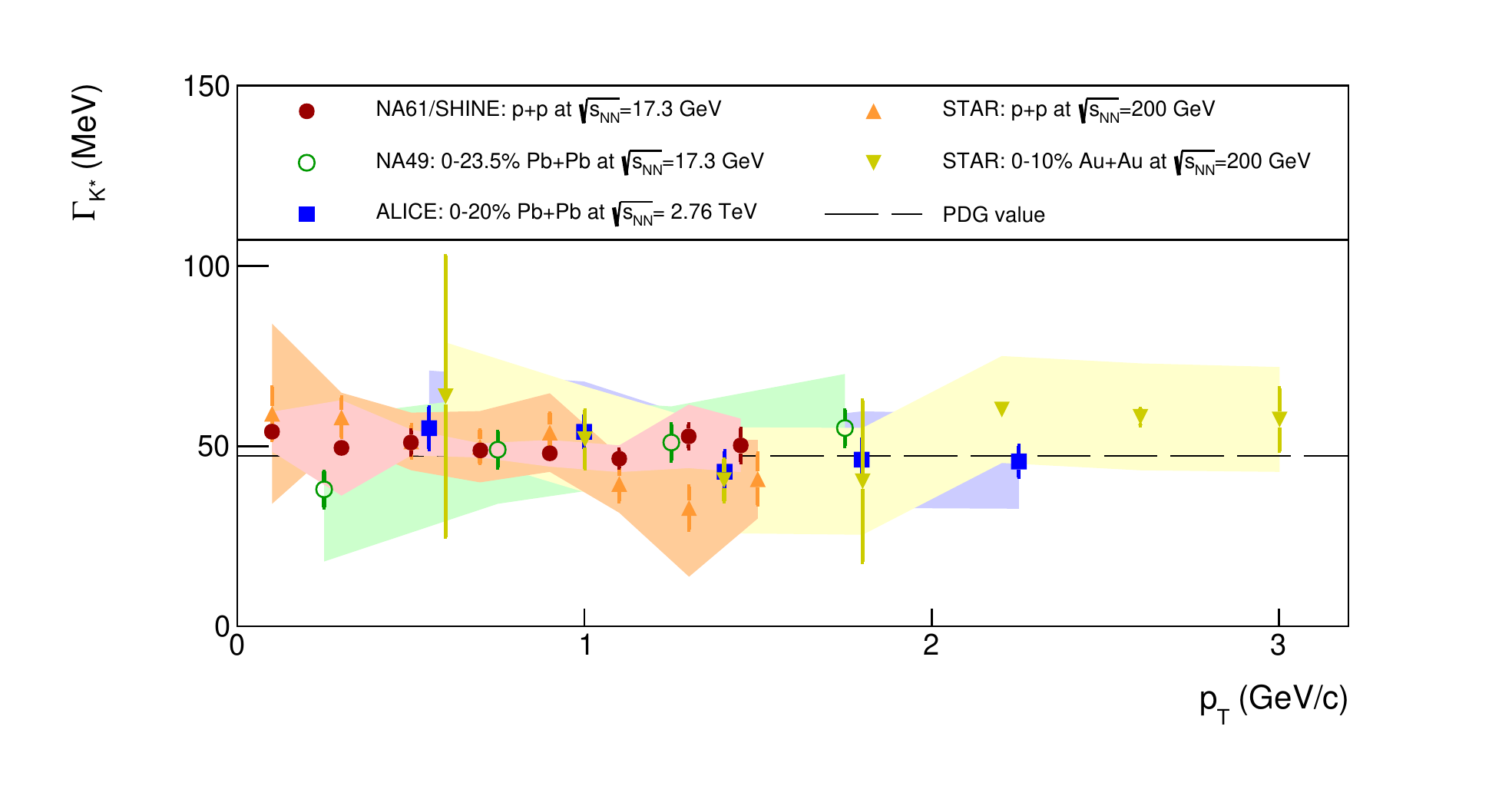}
	\caption{(Color online) The transverse momentum dependence of mass and width of $K^{*}(892)^0$ (or $K^{*0}$) mesons obtained by \NASixtyOne, NA49~\cite{Anticic:2011zr}, ALICE~\cite{Abelev:2014uua} and STAR~\cite{Adams:2004ep}. For ALICE and STAR the averaged ($K^{*0}$) measurements of $K^{*}(892)^0$ and $\overline{K^{*}}(892)^0$ are shown. The horizontal lines represent PDG values~\cite{PDG}. 
 }
	\label{fig:mass_gamma}
\end{figure*}

Figure~\ref{fig:mass_gamma} shows the comparison of mass and width of $K^{*}(892)^0$ mesons obtained in p+p interactions by \NASixtyOne, STAR (top RHIC energy), as well as in Pb+Pb and Au+Au collisions at SPS, RHIC and LHC energies. For the ALICE and STAR experiments the averaged measurements of $K^{*}(892)^0$ and $\overline{K^{*}}(892)^0$ mesons are shown. One sees that among the available results (and within the \pt range covered by the figure)
the precision of the \NASixtyOne measurements is the highest and the results are very close to the PDG values. For p+p collisions the STAR experiment measured lower $K^{*0}$ mass, especially at lower transverse momenta.      

%%%%%%%%%%%%%%%%%%%%%%%%%%%%%%%%%%%%%%%%%%% 

\subsection{Comparison of rapidity spectra and yields with NA49 and \EposLong}

\begin{figure*}
\centering
\includegraphics[width=0.9\textwidth]{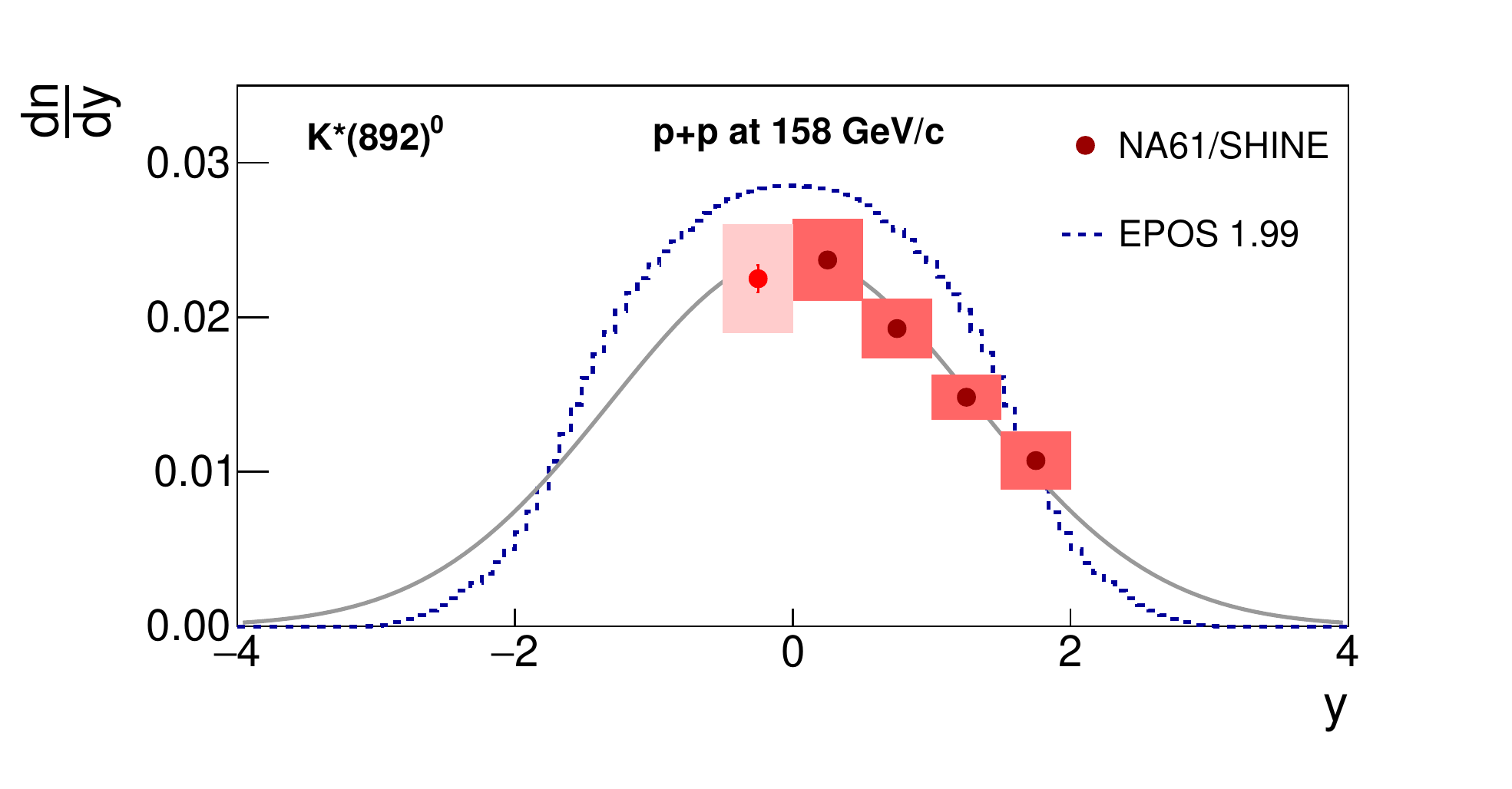}
\vspace{-0.5cm}
\caption[]{(Color online) Comparison of $K^{*}(892)^0$ rapidity distribution from \NASixtyOne (points) and the \EposLong model (dashed line).
The fitted Gaussian function to \NASixtyOne points (solid line) is given by Eq.~(\ref{eq:fit_to_dndy}); the first point (with $y < 0$) was not included in the fit (see the text for details).
}
	\label{fig:dndy_EPOS}
\end{figure*}

The \NASixtyOne measurements of the rapidity spectrum and mean multiplicity were also compared to those predicted by the model of hadron production \EposLong~\cite{Werner:2005jf}. The results are presented in Fig.~\ref{fig:dndy_EPOS} and the numerical values of the multiplicity are listed in Table~\ref{tab:multiplicity}. One sees that the \EposLong model overestimates $K^{*}(892)^0$ production in inelastic p+p collisions at 158~\GeVc.

Table~\ref{tab:multiplicity} also shows the comparison with the NA49 result~\cite{Anticic:2011zr} for the same collision system and beam momentum. 
Instead of analysing in separate \pt bins, as in \NASixtyOne, the NA49 experiment used one wide \pt bin ($0 < p_T < 1.5$~\GeVc). The mean multiplicity of $K^{*}(892)^0$ in NA49 was obtained as the integral under the Gaussian function in the range $-3 < y < 3$ fitted to the $\frac{dn}{dy}$ distribution~\cite{Claudia_H_PHD}. Within the uncertainties shown, the results of both experiments are consistent.

\begin{table}
	\centering
	\begin{tabular}{|c|c|c|}
		\hline
		& $\langle K^{*}(892)^0 \rangle$ & $\sigma_y$ \\
		\hline
		\NASixtyOne, \pt-integrated and extrapolated $\frac{dn}{dy}$ & $(78.44 \pm 0.38 \pm 6.0) \cdot 10^{-3}$ & $1.31 \pm 0.15 \pm 0.09$ \\
		\hline
		NA49, $\frac{dn}{dy}$ in wide \pt bin \cite{Anticic:2011zr} & $(74.1 \pm 1.5 \pm 6.7) \cdot 10^{-3}$ &  $1.17 \pm 0.03 \pm 0.07$ \\
		\hline
		\EposLong, no binning & $(87.82 \pm 0.06) \cdot 10^{-3} $ & - \\
		\hline
	\end{tabular}
	\caption{The mean multiplicities $\langle K^{*}(892)^0 \rangle$ and the widths of the rapidity distributions $\sigma_y$ obtained from $\frac{dn}{dy}$ distributions (see the text for details). The first uncertainty is statistical and the second systematic.}
	\label{tab:multiplicity}
\end{table}

%%%%%%%%%%%%%%%%%%%%%%%%%%%%%%%%%%%%%%%%%%%% 
 
\subsection{System size dependence of $\langle K^{*}(892)^0 \rangle$ at 158\AGeVc and predictions of HGM}

The statistical Hadron Resonance Gas Models (HGM) are commonly used to predict particle multiplicities in elementary and nucleus-nucleus collisions, using as adjustable parameters the chemical freeze-out temperature $T_{chem}$, the baryochemical potential $\mu_B$, strangeness saturation parameter $\gamma_S$, etc. In the following the measured $\langle K^{*}(892)^0 \rangle$ multplicities are compared with predictions of two HGM models described in Refs.~\cite{Becattini:2005xt, Begun:2018qkw}. 

\begin{figure*}
	\centering
	\includegraphics[width=0.7\textwidth]{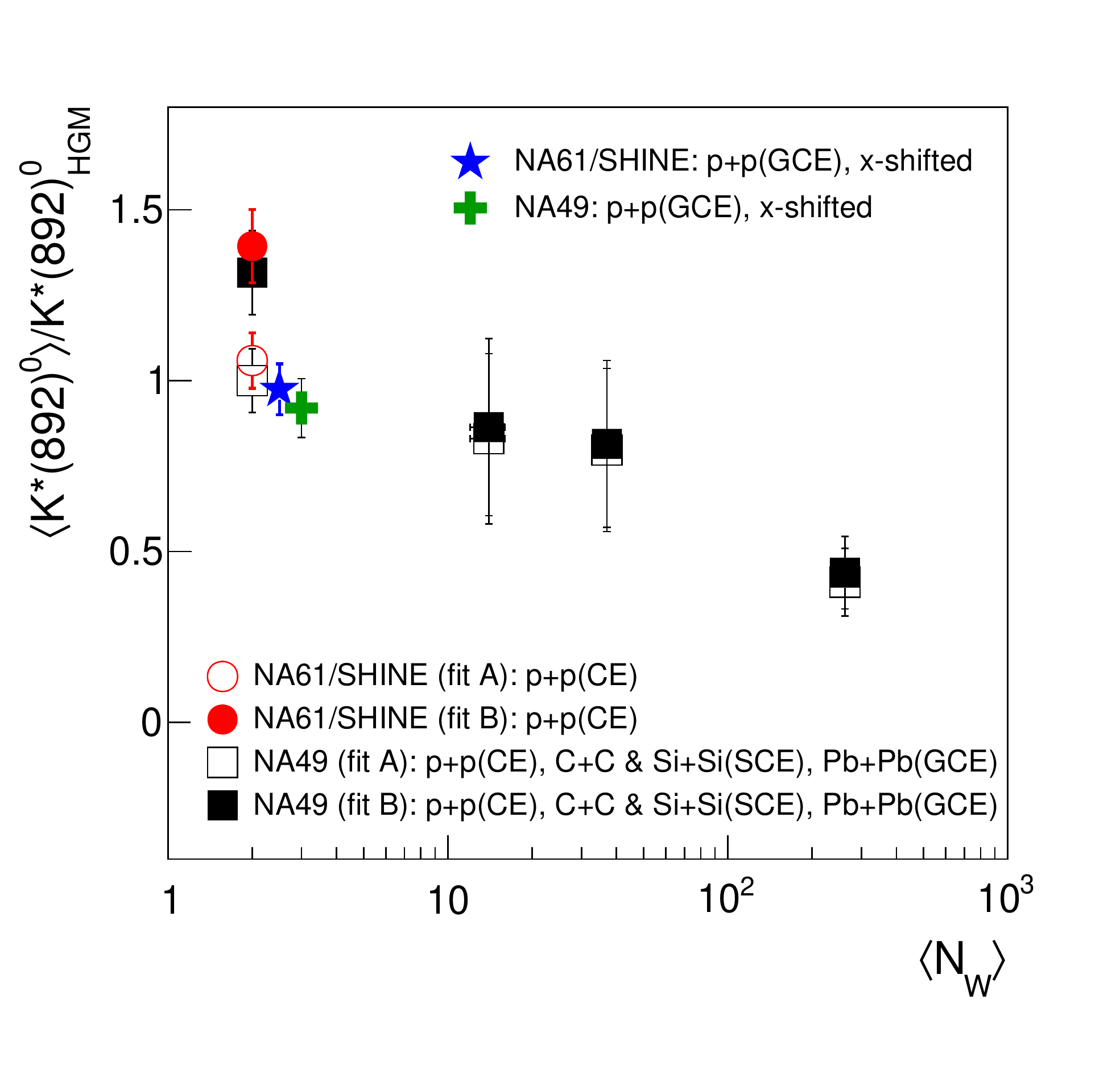}
\vspace{-0.7cm}
	\caption{(Color online) The mean multiplicity of $K^{*}(892)^0$ for p+p reactions (this analysis and NA49 measurement~\cite{Anticic:2011zr}), as well as results of NA49 for C+C, Si+Si and Pb+Pb~\cite{Anticic:2011zr} interactions at 158\AGeVc divided by the HGM predictions~\cite{Becattini:2005xt} for fit B (closed circle and closed squares) and fit A (open circle and open squares), see the text for details. Closed star and cross symbols show p+p measurements compared to HGM predictions for the Grand Canonical Ensemble formulation~\cite{Begun:2018qkw, Begun_priv}. $N_W$ denotes the number of wounded nucleons taken from Ref.~\cite{Anticic:2011zr}. }
	\label{fig:HGM_comb}
\end{figure*}

In Ref.~\cite{Becattini:2005xt} the HGM results for $K^{*}(892)^0$ multiplicities were calculated for two versions of the model fits to particle yields. The first one, called fit B, allowed for strangeness under-saturation so the usual parametrization with $\gamma_S$ was applied. For p+p interactions, the fit was carried out without including the multiplicities of $\Xi$ and $\Omega$ baryons. 
In the second fit, called A, the parameter $\gamma_S$ was replaced by the mean number of strange quark pairs $\langle s \bar{s} \rangle$. For p+p collisions fit A was performed without the $\phi$ meson. For both fits predicted multiplicities were calculated in the Canonical Ensemble (CE)~\cite{Becattini:2005xt}. The measured mean multiplicity of $K^{*}(892)^0$ in 158~\GeVc inelastic p+p interactions was divided by HGM predictions based on fit A and B and compared with the value found by NA49~\cite{Anticic:2011zr}. The results are shown in Fig.~\ref{fig:HGM_comb} for p+p interactions, as well as C+C, Si+Si, and Pb+Pb collisions measured by NA49~\cite{Anticic:2011zr}. In Ref.~\cite{Becattini:2005xt} the S-Canonical Ensemble (SCE) with exact strangeness conservation and grand-canonical treatment of electric charge and baryon number was used for the heavier C+C and Si+Si systems, and the Grand Canonical Ensemble (GCE) was assumed for Pb+Pb collisions. For C+C and Si+Si interactions all available particles were used in the HGM fits, including $\phi$ meson and multi-strange baryons. For Pb+Pb data only the measured $\Lambda(1520)$ yield was removed from the fitted multiplicities. Note that the centrality of Pb+Pb collisions used in the HGM fits was 0--5\% whereas the $\langle K^{*}(892)^0 \rangle$ values in NA49 were obtained for the 0--23.5\% most central interactions. Therefore, the HGM yields had to be scaled by a factor $262/362$ corresponding to the respective number of wounded nucleons (see Table~\ref{tab:Kstar_Kp_Km}).

For heavier systems (including C+C and Si+Si), there is no significant difference between fit A and fit B, however the deviation between the HGM predictions and experimental data increases with increasing system size. The p+p measurements are very close to the HGM prediction but only in case of fit A, where the $\phi$ meson was excluded from the fit. In the most recent paper~\cite{Begun:2018qkw}, where the HGM fits were done for the NA49 and the new \NASixtyOne measurement in p+p interactions, it is also stressed that at SPS energies the $\phi$ meson multiplicities in p+p collisions cannot be well fitted within the CE formulation of the HGM (the quality of CE fits becomes much worse when the $\phi$ meson yield is included).
However, the mean multiplicity of $K^{*}(892)^0$ mesons in inelastic p+p collisions at 158~\GeVc can also be compared to the HGM prediction based on the Grand Canonical Ensemble formulation~\cite{Begun:2018qkw}. The results for the NA49 and \NASixtyOne measurements are shown in Fig.~\ref{fig:HGM_comb} as closed cross and closed star symbols. Surprisingly, the GCE statistical model provides a good description of the $K^{*}(892)^0$ yield in the small p+p system. The numerical values of the \NASixtyOne p+p measurement and the statistical models are presented in the Table~\ref{tab:HGM}. In Fig.~\ref{fig:HGM_comb} the total uncertainty of $\langle K^{*}(892)^0 \rangle$ was taken as the square root of the sum of squares of statistical and systematic uncertainties. The uncertainty of the ratio shown on vertical axis was taken as the final uncertainty of $\langle K^{*}(892)^0 \rangle$ divided by $K^{*}(892)^0_\mathrm{HGM}$.

\begin{table}
	\centering
	\begin{tabular}{|c|c|}
		\hline
		& \multirow{2}{*}{$\langle K^{*}(892)^0 \rangle$ or $K^{*}(892)^0_\mathrm{HGM}$} \\ & \\
		\hline
		\NASixtyOne, \pt-integrated and extrapolated $\frac{dn}{dy}$ & $(78.44 \pm 0.38 \pm 6.0) \cdot 10^{-3}$ \\
		\hline
		HGM, Canonical Ensemble, fit A (no $\phi$) \cite{Becattini:2005xt} & 74.1 $\cdot 10^{-3}$\\
		\hline
		HGM, Canonical Ensemble, fit B (with $\phi$) \cite{Becattini:2005xt} & 56.3 $\cdot 10^{-3}$\\
		\hline
		HGM, Grand Canonical Ensemble (with $\phi$) \cite{Begun:2018qkw, Begun_priv} & 80.5 $\cdot 10^{-3}$ \\
		\hline
	\end{tabular}
	\caption{The mean multiplicity of $K^{*}(892)^0$ mesons for 158~\GeVc inelastic p+p interactions compared to theoretical multiplicities obtained within Hadron Gas Models~\cite{Becattini:2005xt, Begun:2018qkw}.}
	\label{tab:HGM}
\end{table}

%%%%%%%%%%%%%%%%%%%%%%%%%%%%%%%%%%%%%%%%%%%%  

\subsection{$K^*$ over charged kaon ratios and time between freeze-outs}
\label{sec:comparison_time}

The $K^{*}$ to charged kaons ratios may allow to estimate the time interval between chemical and kinetic freeze-out in nucleus-nucleus collisions. The $K^{*}$ mesons have identical quark (anti-quark) content as $K$ mesons, but different mass and relative orientation of quark spins. Thus, the $\langle K^{*}(892)^0 \rangle/ \langle K^{-} \rangle$ and $\langle K^{*}(892)^0 \rangle/ \langle K^{+} \rangle$ ratios are considered as the least model dependent ratios for studying the $K^{*}$ production properties as well as the freeze-out conditions.

\begin{figure*}
	\centering
	\includegraphics[width=0.7\textwidth]{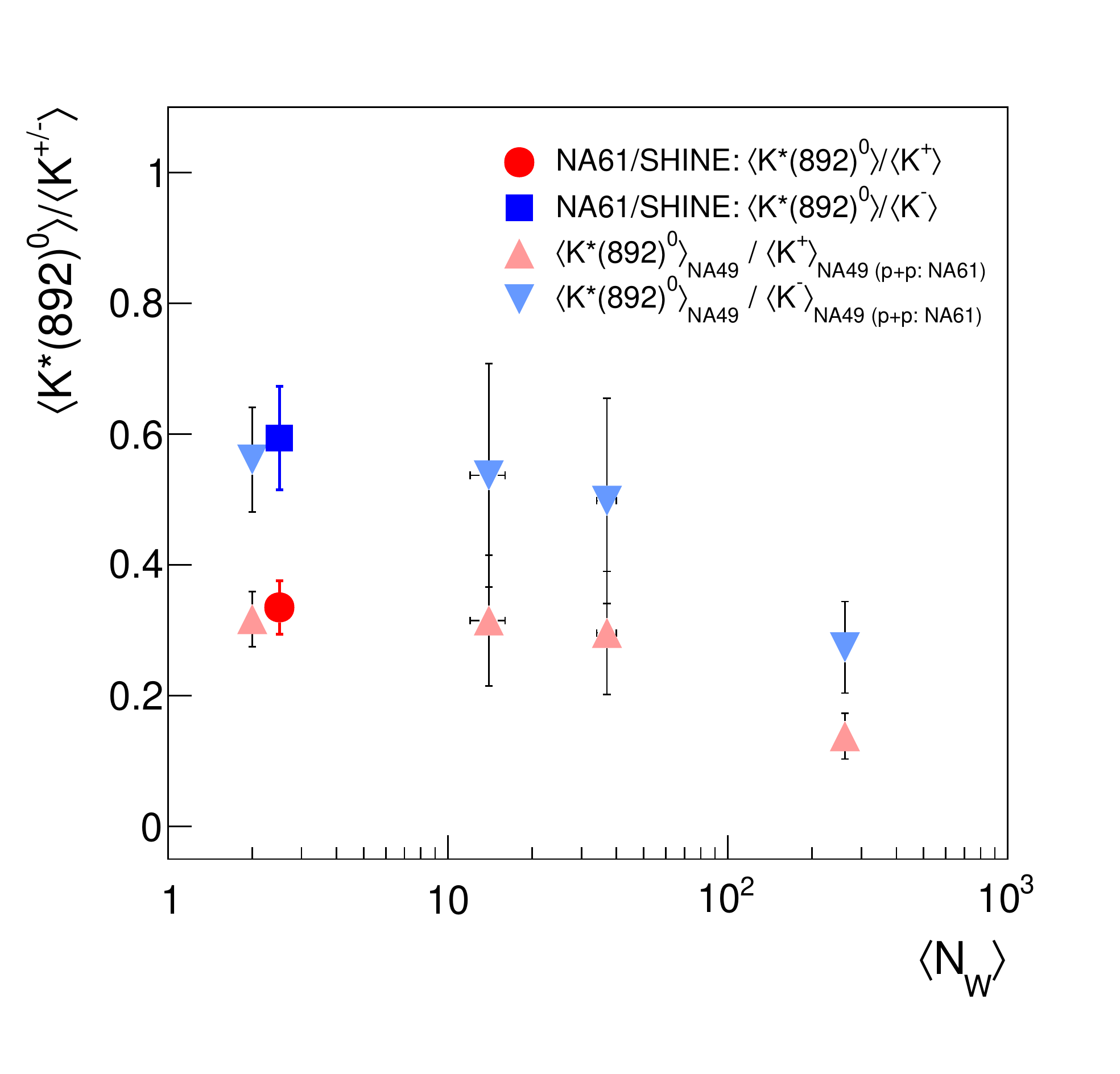}
\vspace{-0.7cm}
	\caption{(Color online) The system size dependences of $\langle K^{*}(892)^0 \rangle/ \langle K^{+} \rangle$ and $\langle K^{*}(892)^0 \rangle/ \langle K^{-} \rangle$ yield ratios in p+p, C+C, Si+Si and Pb+Pb collisions at 158\AGeVc. $N_W$ denotes the number of wounded nucleons taken from Ref.~\cite{Anticic:2011zr}. The numerical values are listed in Table~\ref{tab:Kstar_Kp_Km}. For better visibility the \NASixtyOne points are shifted on the horizontal axis.}
	\label{fig:Kstar_K_ratio}
\end{figure*}

The system size dependence of the $K^{*}/K$ ratio at SPS, RHIC and LHC energies shows a strong decrease with increasing system size and/or multiplicity density (see Sec.~\ref{sec:intro} for a full list of references). The effect seems to be stronger at the SPS than at RHIC and LHC. Figure~\ref{fig:Kstar_K_ratio} presents this dependence at the SPS for the NA49 and \NASixtyOne results at 158\AGeVc. The numerical values are given in Table~\ref{tab:Kstar_Kp_Km}.

\begin{table}
%\small
\footnotesize
	\centering
%\addtolength{\leftskip} {-2cm}
%\addtolength{\rightskip} {-2cm}
	\begin{tabular}{|c|c|c|c|c|c|}
		\hline
		& \multirow{2}{*}{$\langle K^{*}(892)^0 \rangle$} & \multirow{2}{*}{$\langle K^{+} \rangle$} & \multirow{2}{*}{$\langle K^{-} \rangle$} & \multirow{2}{*}{$\langle K^{*}(892)^0 \rangle / \langle K^{+} \rangle$} & \multirow{2}{*}{$\langle K^{*}(892)^0 \rangle / \langle K^{-} \rangle$}  \\ & & & & & \\
		\hline
		\makecell{ \NASixtyOne \\ p+p \\ $N_W$=2} & 0.0784$\pm$0.0060 & 0.234$\pm$0.022 \cite{Aduszkiewicz:2017sei} & 0.132$\pm$0.014 \cite{Aduszkiewicz:2017sei} & 0.335$\pm$0.041 & 0.594$\pm$0.079 \\
		\hline
\makecell{NA49 \\ p+p \\ $N_W$=2} & 0.0741$\pm$0.0069  \cite{Anticic:2011zr} & \makecell{from \\ \NASixtyOne} & \makecell{ from \\ \NASixtyOne } & 0.317$\pm$0.042 & 0.561$\pm$0.080 \\
		\hline
		\makecell{NA49 \\ 15.3\% C+C \\ $\langle N_W \rangle$=14$\pm$2 \cite{Anticic:2011zr}} & 0.80$\pm$0.24 \cite{Anticic:2011zr} & 2.54$\pm$0.25 \cite{Alt:2004wc} & 1.49$\pm$0.16 \cite{Alt:2004wc} & 0.31$\pm$0.10 & 0.54$\pm$0.17 \\
		\hline
		\makecell{NA49 \\ 12.2\% Si+Si \\ $\langle N_W \rangle$=37$\pm$3 \cite{Anticic:2011zr}} & 2.20$\pm$0.66 \cite{Anticic:2011zr} & 7.44$\pm$0.74 \cite{Alt:2004wc} & 4.42$\pm$0.44 \cite{Alt:2004wc} & 0.296$\pm$0.094 & 0.50$\pm$0.16 \\
		\hline
		\makecell{NA49 \\ 23.5\% Pb+Pb \\ $\langle N_W \rangle$=262$\pm$6 \cite{Anticic:2011zr}} & 10.3$\pm$2.5 \cite{Anticic:2011zr} & \makecell{ 74.5$\pm$5.1 \\ (from scaling)} & \makecell{ 37.6$\pm$2.6 \\ (from scaling) }& 0.138$\pm$0.035  & 0.274$\pm$0.070 \\
		\hline
		\makecell{NA49 \\ 5\% Pb+Pb \\ $\langle N_W \rangle$=362$\pm$5 \cite{Afanasiev:2002mx}} & -- & 103.0$\pm$7.1 \cite{Afanasiev:2002mx} & 51.9$\pm$3.6 \cite{Afanasiev:2002mx} & -- & -- \\	
		\hline
	\end{tabular}
	\caption{The mean multiplicities of different particle species measured in nucleus-nucleus collisions at 158\AGeVc by NA49 and \NASixtyOne. The total uncertainties of $\langle K^{*}(892)^0 \rangle$, $\langle K^{+} \rangle$ and $\langle K^{-} \rangle$ were taken as the square roots of the sums of squares of statistical and systematic uncertainties. 
For NA49 p+p data, the $\langle K^{+} \rangle$ and $\langle K^{-} \rangle$ results include statistical uncertainties only ($\langle K^{+} \rangle$ = 0.2267 $\pm$ 0.0006 and $\langle K^{-} \rangle$ = 0.1303 $\pm$ 0.0004), whereas systematic uncertainties for total yields were not reported~\cite{Anticic:2010yg}. Therefore, \NASixtyOne $\langle K^{+} \rangle$ and $\langle K^{-} \rangle$ values were used in the $\langle K^{*}(892)^0 \rangle / \langle K^{+} \rangle$ and $\langle K^{*}(892)^0 \rangle / \langle K^{-} \rangle$ ratios.  
The numbers of $\langle K^{+} \rangle$ and $\langle K^{-} \rangle$ and their uncertainties for the 5\% most central Pb+Pb collisions were multiplied by a factor 262/362 in order to estimate charged kaon multiplicities in the 23.5\% most central Pb+Pb reactions.}
	\label{tab:Kstar_Kp_Km}
\end{table}

The \NASixtyOne $\langle K^{*}(892)^0 \rangle/ \langle K^{+} \rangle$ and $\langle K^{*}(892)^0 \rangle/ \langle K^{-} \rangle$ yield ratios for p+p interactions and the corresponding ratios in central Pb+Pb collisions from NA49 can be used to estimate the time interval between chemical and kinetic freeze-outs in Pb+Pb. Following Ref.~\cite{Adams:2004ep}:

\begin{equation}
\frac{K^{*}}{K} \mid_{kinetic} = \frac{K^{*}}{K} \mid_{chemical} \cdot e^{- \frac{\Delta t}{\tau}},
\label{eq:time_freezeout}
\end{equation}

where: 
	\begin{itemize}
		\item [-] the ratio $\langle K^{*}(892)^0 \rangle/ \langle K^{+/-} \rangle$ in inelastic p+p interactions can be treated as the one at chemical freeze-out,
		\item [-] the ratio $\langle K^{*}(892)^0 \rangle/ \langle K^{+/-} \rangle$ for central Pb+Pb (NA49) interactions can be used as the one at kinetic freeze-out,
		\item [-] $\tau$ is the mean $K^{*}(892)^0$ lifetime of approximately 4.17 fm/c~\cite{PDG},
		\item [-] $\Delta t$ is the time interval between chemical and kinetic freeze-outs calculated in the $K^{*}(892)^0$ rest frame. 
	\end{itemize}

Assuming that the losses of $K^{*}(892)^0$ before kinetic freeze-out are due to rescattering effects and that there are no regeneration processes, the time between chemical and kinetic freeze-outs (in the resonance rest frame) can be estimated as 3.7 $\pm$ 1.2 fm/$c$ from the $\langle K^{*}(892)^0 \rangle/ \langle K^{+} \rangle$ ratio and 3.2 $\pm$ 1.2 fm/$c$ from the $\langle K^{*}(892)^0 \rangle/ \langle K^{-} \rangle$ ratio. These numbers correspond to 23.5\% of the most central Pb+Pb interactions but the time is similar when using 5\% of the most central events. 

Following Ref.~\cite{Acharya:2019qge}, the above times may be expressed in the collision center-of-mass reference system using the multiplicative Lorentz factor:
\begin{equation}
\gamma \approx \sqrt{1+(\langle p_T \rangle /m_0 c)^2},
\end{equation}
where $\langle p_T \rangle$ can be used as an approximation for $K^{*}(892)^0$ total momentum for the measurements at mid-rapidity.
The NA49 experiment published the $K^{*}(892)^0$ transverse momentum spectrum for 23.5\% of the most central Pb+Pb interactions in the rapidity range $0.43 < y < 1.78$ \cite{Anticic:2011zr}. The $\langle p_T \rangle$ can be obtained from the fitted exponential function in the range $0 < \pt < 4$ \GeVc. The average transverse momentum of $K^{*}(892)^0$ mesons was found to be 0.908 \GeVc that results in $\gamma \approx 1.42$. Finally, the Lorentz boosted time interval between chemical and kinetic freeze-outs can be estimated as 5.3 fm/$c$ for the $\langle K^{*}(892)^0 \rangle/ \langle K^{+} \rangle$ ratio or 4.6 fm/$c$ for the $\langle K^{*}(892)^0 \rangle/ \langle K^{-} \rangle$ ratio.

Similar calculations can be performed for the published RHIC ($\sqrt{s_{NN}}=200$ \GeV) and LHC ($\sqrt{s_{NN}}=2760$ \GeV) data. In the STAR experiment at RHIC, the $K^{*0}/K^{-}$ ratio was found to be $0.20 \pm 0.04$ for the 10\% most central Au+Au collisions, and $0.34 \pm 0.05$ for p+p interactions~\cite{Aggarwal:2010mt}. Thus, the time between freeze-outs (calculated in the $K^{*0}$ rest frame) is equal to 2.2 $\pm$ 1.0 fm/$c$. The average transverse momentum of $K^{*0}$ mesons in Au+Au collisions at mid-rapidity ($|y| < 0.5$) was found to be 1.09 \GeVc~\cite{Aggarwal:2010mt}, which corresponds to $\gamma \approx$ 1.57. Therefore, the time interval between freeze-outs, determined in the collision center-of-mass reference system, can be estimated as 3.5 fm/$c$. This value is smaller than the ones obtained at SPS.

In the ALICE experiment at LHC, the $K^{*0}/K^{-}$ ratio was found to be $0.180 \pm 0.027$ for the 5\% most central Pb+Pb collisions, and $0.307 \pm 0.043$ for p+p interactions~\cite{Adam:2017zbf}. Following Eq.~(\ref{eq:time_freezeout}), $\Delta t$ can be evaluated as 2.2 $\pm$ 0.9 fm/$c$. The $\langle p_T \rangle$ of $K^{*0}$ mesons in Pb+Pb collisions at mid-rapidity ($|y| < 0.5$) was found to be 1.310 \GeVc~\cite{Adam:2017zbf}, which corresponds to $\gamma \approx$ 1.77. Thus, at LHC energy, the time interval between freeze-outs, determined in the collision center-of-mass reference system, can be estimated as 3.9 fm/$c$.

The above numbers may imply that, in central heavy ion collisions, the lifetime of the hadronic period of the fireball after chemical freeze-out is longer at SPS than at RHIC or even at LHC energies. 
One should, however, remember that such a conclusion is valid only under the assumption that there are no regeneration processes of $K^{*0}$ mesons before kinetic freeze-out. 
 As the $K^{*}(892)^0$ regeneration may happen at all energies, the obtained time interval values should be considered as lower limits of the time between chemical and kinetic freeze-outs.

%%% Local Variables: 
%%% mode: latex
%%% TeX-master: "main"
%%% End: 

\section{Summary}\label{sec:summary}

In this paper the \NASixtyOne measurement of $K^{*}(892)^0$ meson production via its $K^{+}\pi^{-}$ decay mode in inelastic p+p collisions at beam momentum 158~\GeVc ($\sqrt{s_{NN}}=17.3$~\GeV) was presented. The \textit{template} method was used to extract raw $K^{*}(892)^0$ signals. In this method the background is described as a sum of two contributions: background due to uncorrelated pairs modeled by event mixing and background of correlated pairs modeled by \EposLong.
 For $K^{*}(892)^0$ production the \textit{template} method was found to provide a better background description than the \textit{standard} one which relies on mixed events only. The mass and width of the $K^{*}(892)^0$ were extracted from the fits to background subtracted invariant mass spectra. 
Their values, for different transverse momentum bins, are close to the PDG results, however, a slight increase of the $K^{*}(892)^0$ mass with transverse momentum can be observed. 

With the large statistics of \NASixtyOne data (52.53M events selected by the interaction trigger) it was possible to obtain double-differential transverse momentum and rapidity spectra of $K^{*}(892)^0$ mesons. The full phase-space mean multiplicity of $K^{*}(892)^0$ mesons, obtained from the \pt-integrated and extrapolated rapidity distribution, was found to be $(78.44 \pm 0.38 \pm 6.0) \cdot 10^{-3}$, where the first uncertainty is statistical and the second one is systematic. The result agrees with the previous NA49 measurement for the same system and energy. 

The \NASixtyOne result was compared with predictions of statistical Hadron Resonance Gas models in Canonical and Grand Canonical formulations. Surprisingly, the GCE model provides a good description of the \NASixtyOne measurement of the $K^{*}(892)^0$ multiplicity in p+p collisions. The CE model also agrees provided that the $\phi$ meson is excluded from the fits. 

Finally, based on the previous results of NA49 from central Pb+Pb collisions and the new measurements of \NASixtyOne on p+p interactions, an attempt was made to estimate the time between chemical and kinetic freeze-outs in central Pb+Pb reactions at 158\AGeVc. This time was found to be larger than at RHIC, suggesting that either the system life-time between freeze-outs is indeed higher at SPS or the $K^{*}(892)^0$ regeneration effects start to play a significant role at higher collision energies.

%%% Local Variables: 
%%% mode: latex
%%% TeX-master: "main"
%%% End: 

\section*{Acknowledgements}
We would like to thank the CERN EP, BE, HSE and EN Departments for the
strong support of NA61/SHINE.

This work was supported by
the Hungarian Scientific Research Fund (grant NKFIH 123842\slash123959),
the Polish Ministry of Science
and Higher Education (grants 667\slash N-CERN\slash2010\slash0,
NN\,202\,48\,4339 and NN\,202\,23\,1837), the National Science Centre Poland (grants~2011\slash03\slash N\slash ST2\slash03691,
2013\slash11\slash N\slash ST2\slash03879, 2014\slash13\slash N\slash
ST2\slash02565, 2014\slash14\slash E\slash ST2\slash00018,
2014\slash15\slash B\slash ST2\slash02537 and
2015\slash18\slash M\slash ST2\slash00125, 2015\slash 19\slash N\slash ST2 \slash01689, 2016\slash23\slash B\slash ST2\slash00692, 
2017\slash 25\slash N\slash ST2\slash 02575, 2018\slash 30\slash A\slash ST2\slash 00226),
the Russian Science Foundation, grant 16-12-10176,
the Russian Academy of Science and the
Russian Foundation for Basic Research (grants 08-02-00018, 09-02-00664
and 12-02-91503-CERN), the Ministry of Science and
Education of the Russian Federation, grant No.\ 3.3380.2017\slash4.6,
 the National Research Nuclear
University MEPhI in the framework of the Russian Academic Excellence
Project (contract No.\ 02.a03.21.0005, 27.08.2013),
the Ministry of Education, Culture, Sports,
Science and Tech\-no\-lo\-gy, Japan, Grant-in-Aid for Sci\-en\-ti\-fic
Research (grants 18071005, 19034011, 19740162, 20740160 and 20039012),
the German Research Foundation (grant GA\,1480/8-1), the
Bulgarian Nuclear Regulatory Agency and the Joint Institute for
Nuclear Research, Dubna (bilateral contract No. 4799-1-18\slash 20),
Bulgarian National Science Fund (grant DN08/11), Ministry of Education
and Science of the Republic of Serbia (grant OI171002), Swiss
Nationalfonds Foundation (grant 200020\-117913/1), ETH Research Grant
TH-01\,07-3 
%and the U.S.\ Department of Energy.
and the Fermi National Accelerator Laboratory (Fermilab), a U.S. Department of Energy, Office of Science, HEP User Facility. Fermilab is managed by Fermi Research Alliance, LLC (FRA), acting under Contract No. DE-AC02-07CH11359.

\bibliographystyle{na61Utphys}
\bibliography{main.bbl}

\newpage
{\Large The \NASixtyOne Collaboration}
\bigskip
\begin{sloppypar}
  % based on XML DB with time Thu Nov  7 10:56:03 2019
% Authors in alphabetical order.

\noindent
A.~Aduszkiewicz$^{\,15}$,
E.V.~Andronov$^{\,21}$,
T.~Anti\'ci\'c$^{\,3}$,
V.~Babkin$^{\,19}$,
M.~Baszczyk$^{\,13}$,
S.~Bhosale$^{\,10}$,
A.~Blondel$^{\,23}$,
M.~Bogomilov$^{\,2}$,
A.~Brandin$^{\,20}$,
A.~Bravar$^{\,23}$
% ...

\end{sloppypar}
% based on XML DB with time Thu Nov  7 10:56:03 2019
% Institutes in alphabetical order.

\noindent
% ...
$^{2}$~Faculty of Physics, University of Sofia, Sofia, Bulgaria\\
$^{3}$~Ru{\dj}er Bo\v{s}kovi\'c Institute, Zagreb, Croatia\\
% ...
$^{10}$~Institute of Nuclear Physics, Polish Academy of Sciences, Cracow, Poland\\
% ...
$^{13}$~AGH - University of Science and Technology, Cracow, Poland\\
% ...
$^{15}$~University of Warsaw, Warsaw, Poland\\
% ...
$^{19}$~Joint Institute for Nuclear Research, Dubna, Russia\\
$^{20}$~National Research Nuclear University (Moscow Engineering Physics Institute), Moscow, Russia\\
$^{21}$~St. Petersburg State University, St. Petersburg, Russia\\
% ...
$^{23}$~University of Geneva, Geneva, Switzerland\\
% ...

\end{document}